\input amssym.tex
\input epsf
\epsfclipon


\magnification=\magstephalf
\hsize=14.0 true cm
\vsize=19 true cm
\hoffset=1.0 true cm
\voffset=2.0 true cm

\abovedisplayskip=12pt plus 3pt minus 3pt
\belowdisplayskip=12pt plus 3pt minus 3pt
\parindent=1.0em


\font\sixrm=cmr6
\font\eightrm=cmr8
\font\ninerm=cmr9

\font\sixi=cmmi6
\font\eighti=cmmi8
\font\ninei=cmmi9

\font\sixsy=cmsy6
\font\eightsy=cmsy8
\font\ninesy=cmsy9

\font\sixbf=cmbx6
\font\eightbf=cmbx8
\font\ninebf=cmbx9

\font\eightit=cmti8
\font\nineit=cmti9

\font\eightsl=cmsl8
\font\ninesl=cmsl9

\font\sixss=cmss8 at 8 true pt
\font\sevenss=cmss9 at 9 true pt
\font\eightss=cmss8
\font\niness=cmss9
\font\tenss=cmss10

 at 12 true pt
 at 12 true pt
\font\bigrm=cmr10 at 12 true pt
 at 12 true pt
 at 12 true pt

 at 16 true pt
 at 16 true pt
\font\Bigrm=cmr12 at 16 true pt
 at 16 true pt
 at 16 true pt

\catcode`@=11
\newfam\ssfam

\def\tenpoint{\def\rm{\fam0\tenrm}%
    \textfont0=\tenrm \scriptfont0=\sevenrm \scriptscriptfont0=\fiverm
    \textfont1=\teni  \scriptfont1=\seveni  \scriptscriptfont1=\fivei
    \textfont2=\tensy \scriptfont2=\sevensy \scriptscriptfont2=\fivesy
    \textfont3=\tenex \scriptfont3=\tenex   \scriptscriptfont3=\tenex
    \textfont\itfam=\tenit                  \def\it{\fam\itfam\tenit}%
    \textfont\slfam=\tensl                  \def\sl{\fam\slfam\tensl}%
    \textfont\bffam=\tenbf \scriptfont\bffam=\sevenbf
    \scriptscriptfont\bffam=\fivebf
                                            \def\bf{\fam\bffam\tenbf}%
    \textfont\ssfam=\tenss \scriptfont\ssfam=\sevenss
    \scriptscriptfont\ssfam=\sevenss
                                            \def\ss{\fam\ssfam\tenss}%
    \normalbaselineskip=13pt
    \setbox\strutbox=\hbox{\vrule height8.5pt depth3.5pt width0pt}%
    \let\big=\tenbig
    \normalbaselines\rm}

\def\ninepoint{\def\rm{\fam0\ninerm}%
    \textfont0=\ninerm      \scriptfont0=\sixrm
                            \scriptscriptfont0=\fiverm
    \textfont1=\ninei       \scriptfont1=\sixi
                            \scriptscriptfont1=\fivei
    \textfont2=\ninesy      \scriptfont2=\sixsy
                            \scriptscriptfont2=\fivesy
    \textfont3=\tenex       \scriptfont3=\tenex
                            \scriptscriptfont3=\tenex
    \textfont\itfam=\nineit \def\it{\fam\itfam\nineit}%
    \textfont\slfam=\ninesl \def\sl{\fam\slfam\ninesl}%
    \textfont\bffam=\ninebf \scriptfont\bffam=\sixbf
                            \scriptscriptfont\bffam=\fivebf
                            \def\bf{\fam\bffam\ninebf}%
    \textfont\ssfam=\niness \scriptfont\ssfam=\sixss
                            \scriptscriptfont\ssfam=\sixss
                            \def\ss{\fam\ssfam\niness}%
    \normalbaselineskip=12pt
    \setbox\strutbox=\hbox{\vrule height8.0pt depth3.0pt width0pt}%
    \let\big=\ninebig
    \normalbaselines\rm}

\def\eightpoint{\def\rm{\fam0\eightrm}%
    \textfont0=\eightrm      \scriptfont0=\sixrm
                             \scriptscriptfont0=\fiverm
    \textfont1=\eighti       \scriptfont1=\sixi
                             \scriptscriptfont1=\fivei
    \textfont2=\eightsy      \scriptfont2=\sixsy
                             \scriptscriptfont2=\fivesy
    \textfont3=\tenex        \scriptfont3=\tenex
                             \scriptscriptfont3=\tenex
    \textfont\itfam=\eightit \def\it{\fam\itfam\eightit}%
    \textfont\slfam=\eightsl \def\sl{\fam\slfam\eightsl}%
    \textfont\bffam=\eightbf \scriptfont\bffam=\sixbf
                             \scriptscriptfont\bffam=\fivebf
                             \def\bf{\fam\bffam\eightbf}%
    \textfont\ssfam=\eightss \scriptfont\ssfam=\sixss
                             \scriptscriptfont\ssfam=\sixss
                             \def\ss{\fam\ssfam\eightss}%
    \normalbaselineskip=10pt
    \setbox\strutbox=\hbox{\vrule height7.0pt depth2.0pt width0pt}%
    \let\big=\eightbig
    \normalbaselines\rm}

\def\tenbig#1{{\hbox{$\left#1\vbox to8.5pt{}\right.\n@space$}}}
\def\ninebig#1{{\hbox{$\textfont0=\tenrm\textfont2=\tensy
                       \left#1\vbox to7.25pt{}\right.\n@space$}}}
\def\eightbig#1{{\hbox{$\textfont0=\ninerm\textfont2=\ninesy
                       \left#1\vbox to6.5pt{}\right.\n@space$}}}

\font\sectionfont=cmbx10
\font\subsectionfont=cmti10

\def\figurecaptionfont{\ninepoint}
\def\tablecaptionfont{\ninepoint}
\def\footnotefont{\eightpoint}


\newcount\equationno
\newcount\bibitemno
\newcount\figureno
\newcount\tableno

\equationno=0
\bibitemno=0
\figureno=0
\tableno=0


\footline={\ifnum\pageno=0{\hfil}\else
{\hss\rm\the\pageno\hss}\fi}


\def\section #1. #2 \par
{\vskip0pt plus .10\vsize\penalty-100 \vskip0pt plus-.10\vsize
\vskip 1.6 true cm plus 0.2 true cm minus 0.2 true cm
\global\def\equationlabel{#1}
\global\equationno=0
\leftline{\sectionfont #1. #2}\par
\immediate\write\terminal{Section #1. #2}
\vskip 0.7 true cm plus 0.1 true cm minus 0.1 true cm
\noindent}


\def\subsection #1 \par
{\vskip0pt plus 0.8 true cm\penalty-50 \vskip0pt plus-0.8 true cm
\vskip2.5ex plus 0.1ex minus 0.1ex
\leftline{\subsectionfont #1}\par
\immediate\write\terminal{Subsection #1}
\vskip1.0ex plus 0.1ex minus 0.1ex
\noindent}


\def\appendix #1. #2 \par
{\vskip0pt plus .20\vsize\penalty-100 \vskip0pt plus-.20\vsize
\vskip 1.6 true cm plus 0.2 true cm minus 0.2 true cm
\global\def\equationlabel{\hbox{\rm#1}}
\global\equationno=0
\leftline{\sectionfont Appendix #1. #2}\par
\immediate\write\terminal{Appendix #1. #2}
\vskip 0.7 true cm plus 0.1 true cm minus 0.1 true cm
\noindent}



\def\equation#1{$$\displaylines{\qquad #1}$$}
\def\enum{\global\advance\equationno by 1
\hfill\llap{{\rm(\equationlabel.\the\equationno)}}}
\def\noenum{\hfill}
\def\next#1{\cr\noalign{\vskip#1}\qquad}


\def\ifundefined#1{\expandafter\ifx\csname#1\endcsname\relax}

\def\ref#1{\ifundefined{#1}?\immediate\write\terminal{unknown reference
on page \the\pageno}\else\csname#1\endcsname\fi}

\newwrite\terminal
\newwrite\bibitemlist

\def\bibitem#1#2\par{\global\advance\bibitemno by 1
\immediate\write\bibitemlist{\string\def
\expandafter\string\csname#1\endcsname
{\the\bibitemno}}
\item{[\the\bibitemno]}#2\par}

\def\beginbibliography{
\vskip0pt plus .15\vsize\penalty-100 \vskip0pt plus-.15\vsize
\vskip 1.2 true cm plus 0.2 true cm minus 0.2 true cm
\leftline{\sectionfont References}\par
\immediate\write\terminal{References}
\immediate\openout\bibitemlist=biblist
\frenchspacing\parindent=1.8em
\vskip 0.5 true cm plus 0.1 true cm minus 0.1 true cm}

\def\endbibliography{
\immediate\closeout\bibitemlist
\nonfrenchspacing\parindent=1.0em}


\def\figurecaption#1{\global\advance\figureno by 1
\narrower\figurecaptionfont
Fig.~\the\figureno. #1}

\def\tablecaption#1{\global\advance\tableno by 1
\vbox to 0.25 true cm { }
\centerline{\tablecaptionfont%
Table~\the\tableno. #1}
\vskip-0.4 true cm}

\def\thicktablerule{\hrule height1pt}
\def\thintablerule{\hrule height0.4pt}

\tenpoint

\immediate\openin\bibitemlist=biblist
\ifeof\bibitemlist\immediate\closein\bibitemlist
\else\immediate\closein\bibitemlist
\input biblist \fi


\def\thismonth{\ifcase\month\or
January\or February\or March\or April\or May\or June\or
July\or August\or September\or October\or November\or December\fi}



\def\rmd{{\rm d}}

\def\rme{{\rm e}}
\def\rmO{{\rm O}}


\def\Re{{\rm Re}\,}


\def\proof{\noindent{\sl Proof:}\kern0.6em}

\def\frac#1#2{\hbox{$#1\over#2$}}
\def\dual{\mathstrut^*\kern-0.1em}

\def\lvec#1{\setbox0=\hbox{$#1$}
    \setbox1=\hbox{$\scriptstyle\leftarrow$}
    #1\kern-\wd0\smash{
    \raise\ht0\hbox{$\raise1pt\hbox{$\scriptstyle\leftarrow$}$}}
    \kern-\wd1\kern\wd0}
\def\rvec#1{\setbox0=\hbox{$#1$}
    \setbox1=\hbox{$\scriptstyle\rightarrow$}
    #1\kern-\wd0\smash{
    \raise\ht0\hbox{$\raise1pt\hbox{$\scriptstyle\rightarrow$}$}}
    \kern-\wd1\kern\wd0}
\def\slash#1{\setbox0=\hbox{$#1$}\setbox1=\hbox{$\kern1pt/$}
    #1\kern-\wd0\kern1pt/\kern-\wd1\kern\wd0}


\def\nabstar#1{{\nabla\kern0.5pt\smash{\raise 4.5pt\hbox{$\ast$}}
               \kern-5.5pt_{#1}}}

\def\drvstar#1{{\partial\kern0.5pt\smash{\raise 4.5pt\hbox{$\ast$}}
               \kern-6.0pt_{#1}}}

\def\ldrvstar#1{{\lvec{\,\partial}\kern-0.5pt\smash{\raise 4.5pt\hbox{$\ast$}}
               \kern-5.0pt_{#1}}}


\def\MeV{{\rm MeV}}

\def\fm{{\rm fm}}



\def\psibar{\overline{\psi}}


\def\dirac#1{\gamma_{#1}}
\def\diracstar#1#2{
    \setbox0=\hbox{$\gamma$}\setbox1=\hbox{$\gamma_{#1}$}
    \gamma_{#1}\kern-\wd1\kern\wd0
    \smash{\raise4.5pt\hbox{$\scriptstyle#2$}}}


\def\SUtwo{{\rm SU(2)}}
\def\SUthree{{\rm SU(3)}}

\def\tr{{\rm tr}}

\def\Ad{{\rm Ad}\kern0.1em}
\def\Exp{{\rm E}}


\def\B{\Lambda}
\def\Bs{\B^{\kern-0.1em*}}
\def\dB{\partial\B}

\def\DB{D_{\B}}

\def\DdB{D_{\dB}}

\def\DhatB{\hat{D}_{\B}}

\def\Om{\Omega}
\def\Oms{\Omega^{\ast}}
\def\dOm{\partial\Om}
\def\dOms{\partial\Oms}
\def\dOmsring{[\dOms]}
\def\DOm{D_{\Om}}
\def\DOms{D_{\Oms_{\vphantom{1}}}}
\def\DdOm{D_{\dOm}}
\def\DdOms{D_{\dOms_{\vphantom{1}}}}

\def\PdOms{P_{\dOms}}
\def\VdOms{V_{\dOms}}
\def\tOm{\theta_{\Om}}
\def\tOms{\theta_{\Oms}}


\def\Nkv{\langle N_{\rm GCR}\rangle}
\def\Ntr{N_{\rm tr}}
\def\tauint{\tau_{\rm int}}
\def\Dw{D_{\rm w}}
\def\Dhat{\hat{D}}
\def\SG{S_{\rm G}}

\def\Tup{{\frak T}}
\def\Int{{\frak I}}
\def\step{\epsilon}
\def\Pacc{P_{\rm acc}}
\def\resF{r}
\def\resS{\tilde{r}}
\def\tint{\tau_{\rm int}}
\def\abar{\bar{a}}
\def\Gbar{\overline{\Gamma}}
\def\rhobar{\bar{\rho}}
\def\fpp{f_{\rm PP}}
\def\fap{f_{\rm AP}}
\def\mpi{m_{\pi}}

%
\rightline{CERN-PH-TH/2004-177}

\vskip 1.5cm 
\centerline{\Bigrm Schwarz-preconditioned HMC algorithm}
\vskip 0.3 true cm
\centerline{\Bigrm for two-flavour lattice QCD}
\vskip 0.6 true cm
\centerline{\bigrm Martin L\"uscher}
\vskip1ex
\centerline{\it CERN, Physics Department, TH Division}
\centerline{\it CH-1211 Geneva 23, Switzerland}
\vskip 0.8 true cm
\thintablerule
\vskip 2.0ex
\ninepoint
\leftline{\bf Abstract}
\vskip 1.0ex\noindent
The combination of a non-overlapping Schwarz preconditioner and 
the Hybrid Monte Carlo (HMC) algorithm is shown to 
yield an efficient simulation algorithm for two-flavour
lattice QCD with Wilson quarks.
Extensive tests are performed, on lattices of size up to
$32\times24^3$, with lattice spacings $a\simeq0.08\,\fm$ and at 
bare current-quark masses as low as
$21\,\MeV$.
\vskip 2.0ex
\thintablerule

\tenpoint

\vskip-0.2cm

\section 1. Introduction

At present, perhaps the greatest obstacle in lattice QCD is the fact that
the efficiency of the established simulation algorithms 
rapidly decreases when the continuum limit is approached and
the masses of the light quarks are scaled towards their physical values
[\ref{SesamTXLauto}--\ref{CPPACSsmall}].
The dynamics of these algorithms is still not fully understood,
but it is quite clear that the poor scaling behaviour is driven by
the condition number of the lattice Dirac operator, which grows 
inversely proportionally to the lattice spacing 
and the quark mass.

Preconditioning is usually perceived as a technique 
for the efficient solution 
of ill-conditioned systems of linear equations [\ref{Saad}]. 
This kind of preconditioning is routinely applied
in lattice QCD to accelerate the solver for the lattice Dirac equation.
While the solver is a central element of
the HMC simulation algorithm [\ref{HMC}],
it is also possible to precondition this algorithm
itself, using another preconditioner perhaps, 
by factorizing the quark determinant
into the determinants of the preconditioners and
the preconditioned Dirac operator.
The magnitude of the quark force terms in 
the molecular dynamics equations is then often reduced,
which allows the associated integration step sizes
to be set to larger values and thus
leads to an acceleration of the algorithm
[\ref{ForcrandTakaishi}--\ref{DellaMorteEtAl},\ref{JLQCDlight}].

In the present paper the effectiveness of 
a non-overlapping Schwarz precondi\-tio\-ner
for the HMC algorithm is studied.
The application of the Schwarz alternating procedure in lattice
QCD has previously been advocated 
in refs.~[\ref{SchwarzAlgorithm},\ref{SchwarzSolver}], and some 
familiarity with the second of these papers will be assumed here.
In order to bring out the underlying strategies more clearly,
the general structure of the preconditioned
HMC algorithm is first discussed.
The Schwarz-preconditioned algorithm
is then introduced in sect.~3 and 
the results of some test runs are reported in sect.~4.

\section 2. Preconditioned HMC algorithm

Only the standard Wilson formulation [\ref{Wilson}] of lattice QCD 
will be considered in this paper,
with bare coupling $g_0$ and bare quark mass $m_0$,
but the algorithm is set up in such a way that 
$\rmO(a)$ improvement [\ref{SW},\ref{OaImp}] 
and more complicated gauge actions with double-plaquette terms
[\ref{Iwasaki},\ref{LWaction}]
can be easily included. 
As usual a four-dimensional hypercubic lattice
of size $T\times L^3$ will be assumed,
with lattice spacing $a$ set to unity for convenience, and
periodic boundary conditions in all directions,
except for the quark fields, which are taken to be antiperiodic in time.
Any unexplained notations are 
as in ref.~[\ref{SchwarzSolver}].

\subsection 2.1 Factorization of the quark determinant

In the familiar case of even--odd preconditioning,
the lattice points are ordered such that the even ones come first
and the (massive) Wilson--Dirac operator $D\equiv\Dw+m_0$
then assumes the block form
\equation{
   D=\pmatrix{A_{11} & A_{12} \cr A_{21} & A_{22}\cr}
   \enum
}
in position space.
Whenever the Dirac operator is written in this way,
its determinant may be factorized according to
\equation{
   \det D=\det A_{11} \det A_{22} \det\left\{1-A_{11}^{-1}A_{12}
   A_{22}^{-1}A_{21}\right\},
   \enum
}
where the operator in the curly bracket is referred to 
as the preconditioned Dirac operator or (in the mathematical literature)
as the {\it Schur complement}\/ of the Dirac operator with respect to the 
block decomposition (2.1).

In general, preconditioning is always associated with
a factorization
\equation{
   \det D=\det R_1\ldots\det R_n
   \enum
}
of the quark determinant into the determinants of certain operators $R_k$. 
How many factors one obtains,
and in exactly which spaces of
fields the operators act, depends on 
the chosen preconditioner. It is also possible to combine
several preconditioners, which may result in further factorizations.
Just to mention an example,
the simple polynomial preconditioning of 
the even--odd preconditioned Wilson--Dirac operator $\Dhat$, 
which was recently considered in 
refs.~[\ref{Hasenbusch}--\ref{DellaMorteEtAl}], 
leads to
\equation{
   R_1=\Dhat+M,\qquad R_2=\Dhat(\Dhat+M)^{-1},
   \enum
}
where $M$ is an adjustable mass parameter.

\subsection 2.2 Preconditioned molecular dynamics

For any given factorization of the quark determinant,
the HMC algorithm for two-flavour QCD can be set up in the standard manner.
First the link momenta $\Pi(x,\mu)$ and the appropriate
pseudo-fermion fields $\phi_1(x),\ldots,\phi_n(x)$
(one for each factor of the determinant)
need to be introduced. These fields are elements of certain linear spaces that
are assumed to be equipped
with the obvious scalar products.
The HMC hamiltonian may then be
written in the compact form
\equation{
   H=\frac{1}{2}\left(\Pi,\Pi\right)+\SG+
   \sum_{k=1}^n\left(R_k^{-1}\phi_k,R_k^{-1}\phi_k\right),
   \enum
}
where $\SG$ denotes the action of the gauge field.

Starting from this expression,
the forces in the molecular dynamics equations
\equation{
   {\rmd\over\rmd\tau}\Pi(x,\mu)=-\sum_{k=0}^nF_k(x,\mu),
   \enum
   \next{1ex}
   {\rmd\over\rmd\tau}U(x,\mu)=\Pi(x,\mu)U(x,\mu),
   \enum
}
are obtained by differentiation with
respect to the link variables. They take values
in the Lie algebra of $\SUthree$ and 
are such that
\equation{
   (\omega,F_0)=\delta_{\omega}\SG,
   \enum
   \next{1.5ex}
   (\omega,F_k)=2\kern1pt\Re\kern-2.5pt
   \left(R_k^{-1}\phi_k,\delta_{\omega}R_k^{-1}\phi_k\right),
   \qquad k=1,\ldots,n,
   \enum
}
for all infinitesimal variations
\equation{
   \delta_{\omega}U(x,\mu)=\omega(x,\mu)U(x,\mu)
   \enum
}
of the gauge field.

Among the operators $R_k$ there is normally one, say $R_n$, 
which is the preconditioned Wilson--Dirac operator or 
a closely related operator.
In the course of the numerical integration of the molecular
dynamics equations, the linear equation
\equation{
   R_n\psi=\eta
   \enum
}
must be solved many times and this part of the calculation
is then likely to consume most of the computer time, particularly
at small quark masses.
An important point to note here is that the solution of eq.~(2.11)
can be obtained by solving the equivalent 
Wilson--Dirac equation, using
a pre\-con\-ditioner and a solver that are optimal for this task.
In the case of the factorization (2.2), for example,
there are $n=3$ operators,
\equation{
  R_1=A_{11},\quad R_2=A_{22},\quad R_3=1-A_{11}^{-1}A_{12}
  A_{22}^{-1}A_{21},
  \enum
}
and eq.~(2.11) is equivalent to
\equation{
  D\chi=\pmatrix{A_{11}\eta \cr 0 \cr},
  \qquad \chi=\pmatrix{\psi \cr \ast \cr}.
  \enum
}
The preconditioner used for the solution of this equation
and the one that determines the factorization of the 
quark determinant thus do not need to be the same.
These two kinds of preconditioning should be kept separate and
must actually satisfy quite different criteria, as will become clear below.

\subsection 2.3 Multiple step size integration

The widely used 
numerical integration schemes for the molecular dynamics equations
can be considered to be sequences of elementary steps
in which either all momenta $\Pi(x,\mu)$ or all link variables $U(x,\mu)$ are
updated. For a step of size $\step$, 
these elementary operations are 
\equation{
   \Tup_k(\step):\kern1.5pt\quad\Pi(x,\mu)\to\Pi(x,\mu)-\step F_k(x,\mu),
   \qquad k=0,\ldots,n,
   \enum
   \next{2ex}
   \Tup_{U}(\step):\quad U(x,\mu)\to\Exp\left(\step\Pi(x,\mu)\right)\kern-0.5pt
   U(x,\mu),
   \enum
}
where $\Exp(X)$ stands for the 
SU(3) exponential function or some suitable approximation to it
(see appendix A).
The leap-frog integration from molecular dynamics
time $0$ to $\tau$, for example,
amounts to the application of the product
\equation{
   \left\{
   \Tup_0(\frac{1}{2}\step)\ldots\Tup_n(\frac{1}{2}\step)
   \Tup_{U}(\step)
   \Tup_0(\frac{1}{2}\step)\ldots\Tup_n(\frac{1}{2}\step)
   \right\}^N, 
   \qquad \step={\tau\over N},
   \enum
}
to the initial field configuration. 
As is well known,
this popular integrator converges  
with a rate proportional to $\step^2$
and satisfies the basic theoretical requirements, 
i.e.~it preserves the measure in phase space and 
is exactly time-reversible.

If the forces $F_0,\ldots,F_n$ have sizeably different magnitudes
such that, say, $\|F_0\|$ is larger than $\|F_1\|$, 
which is larger than $\|F_2\|$, and so on,
it is possible to accelerate the integration  
by adopting a hierarchical scheme
[\ref{SextonWeingarten},\ref{PeardonSexton}]. 
First an integrator 
\equation{
  \Int_{0}(\tau,N_0)= 
  \left\{
  \Tup_0(\frac{1}{2}\step)
  \Tup_{U}(\step)
  \Tup_0(\frac{1}{2}\step)
  \right\}^{N_0}, 
  \qquad \step={\tau\over N_0},
  \enum
}
is defined, which neglects all quark forces $F_1,\ldots,F_n$.
More complicated integrators that include an increasing number of 
these forces are then constructed recursively through
\equation{
  \Int_k(\tau,N_0,\ldots,N_k)=
  \noenum
  \next{1ex}
  \kern2.5em\left\{
  \Tup_k(\frac{1}{2}\step)
  \Int_{k-1}(\step,N_0,\ldots,N_{k-1})
  \Tup_k(\frac{1}{2}\step)
  \right\}^{N_k}, 
  \qquad \step={\tau\over N_k}.
  \enum
}
The full integrator $\Int_n(\tau,N_0,\ldots,N_n)$ that is 
obtained in this way is characterized by the
time intervals
\equation{
  \step_k={\tau\over N_kN_{k+1}\ldots N_n}
  \enum
}
at which the forces $F_k$ need to be evaluated. 

In practice the step numbers $N_k$ must be such
that the shifts $\step_0\Pi$ and $\step_kF_k$
remain small along the molecular dynamics trajectory.
The smaller the forces are the larger the step sizes can be,
and whether a reduction of the computational effort 
is achieved thus depends on the magnitudes of the forces and the difficulty
to compute them.

\subsection 2.4 Choosing a preconditioner

In general the preconditioning of the HMC algorithm is beneficial because
it leads to smaller quark forces along the molecular dynamics trajectories.
Ideally the preconditioning should be such that the forces
that are relatively expensive to calculate (in terms of the required computer
time) are those with the smallest magnitude. 

Preconditioning may also soften the scaling behaviour
of the simulation algorithm with respect to the quark mass.
When the chiral limit is approached,
the quark forces tend to increase, and probably become more
irregular too, which requires the step sizes to be adjusted accordingly.
In a recent study of two-flavour QCD at small quark masses
by Namekawa et al.~[\ref{CPPACSsmall}], for example,
the step sizes had to be decreased roughly in proportion to the 
quark mass. The choice of the preconditioner can have an
influence on this behaviour, as will be shown later in this paper.

Eventually the performance of the preconditioned algorithm
must be determined empirically, since the dependence
of the autocorrelation times of the quantities of interest
on the preconditioner is difficult to foresee.
The suitability of the 
algorithm for parallel processing and other 
practical issues
may also have to be taken into account at this point.

\section 3. Schwarz-preconditioned algorithm

The classical Schwarz alternating procedure
was previously considered as 
a preconditioner for the Wilson--Dirac equation
[\ref{SchwarzSolver}].
When combined with the 
generalized conjugate residual (GCR) algorithm,
this preconditioner
proved to be very efficient, and it 
will now be used, in its simplest form with cycle number $n_{\rm cy}=1$,
to precondition the HMC algorithm.

\subsection 3.1 Domain decomposition

Following ref.~[\ref{SchwarzSolver}]
the lattice is covered by a regular grid of non-overlapping
rectangular blocks $\B$.
For technical reasons
the block sizes are assumed to be even and to be such
that the blocks can be 
chessboard-coloured, i.e.~that there is an even number of blocks
in all dimensions. The union of the black blocks is then denoted
by $\Om$ and the union of the white blocks by $\Oms$
(see fig.~1).

\topinsert
\vbox{
\vskip0.0cm
\epsfxsize=7.0cm\hskip2.5cm\epsfbox{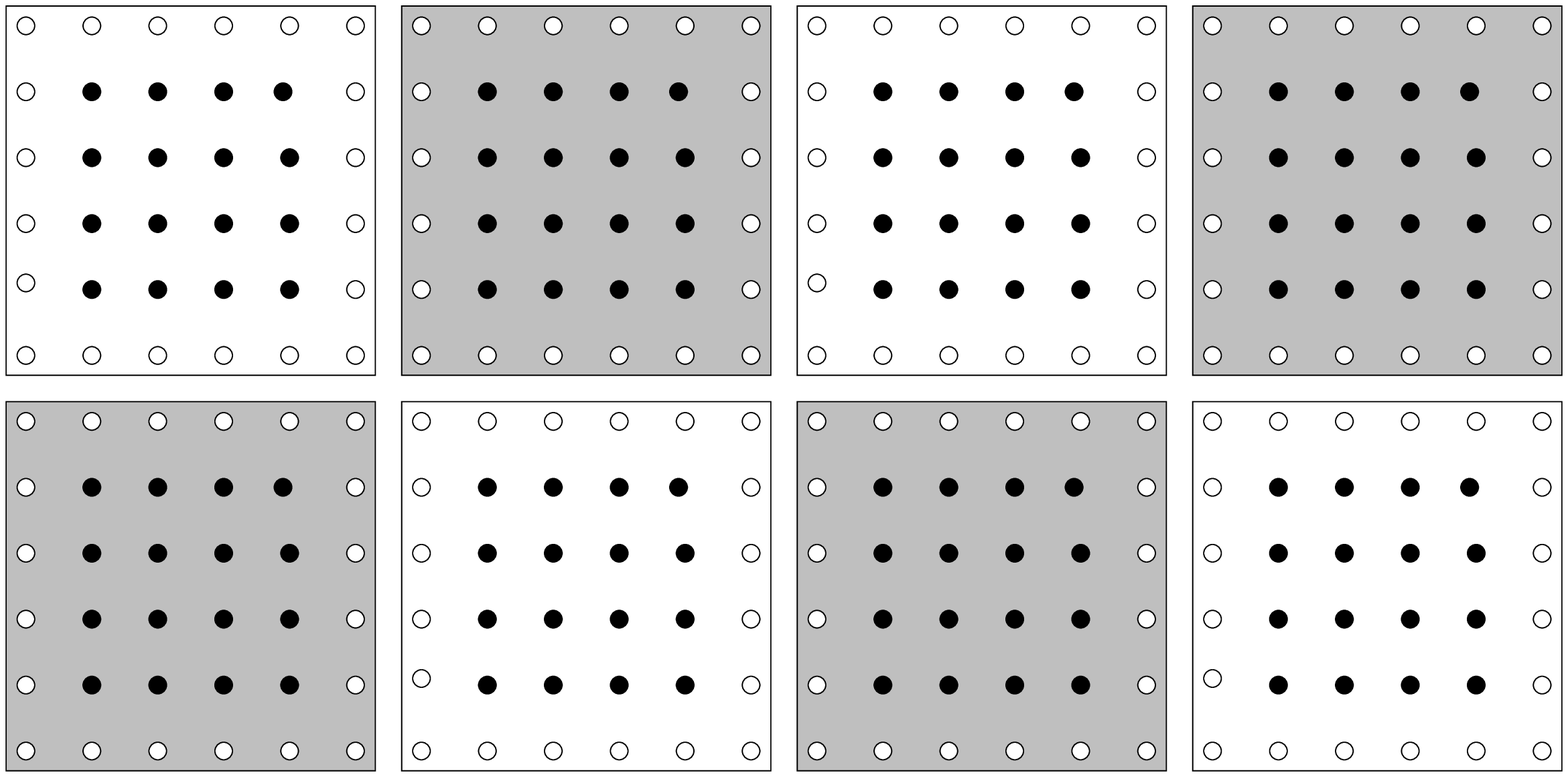}
\vskip0.3cm
\figurecaption{%
Two-dimensional cross-section of 
a $24\times12^3$ lattice covered by non-overlap\-ping $6^4$ blocks $\B$.
The domains $\Om$ and $\Oms$ are the unions of the black and white blocks
respectively, and their exterior boundaries $\dOm$ and $\dOms$ consist 
of all points in the 
complementary domain represented by open circles.
}
\vskip0.0cm
}
\endinsert

With respect to an ordering of the lattice points where those 
in $\Om$ come first, the Wilson--Dirac operator assumes the block
form
\equation{
  D=\pmatrix{\DOm &\DdOm \cr \DdOms &\DOms\cr}.
  \enum
}
The operator $\DOm$, for example, 
coincides with the Wilson--Dirac operator on $\Om$ with Dirichlet
boundary conditions, while $\DdOm$ is the sum of 
all hopping terms from the exterior boundary $\dOm$ of $\Om$ to 
the boundary $\dOms$ of $\Oms$.

It is often convenient to let these operators act on
quark fields that are defined on the whole lattice  
rather than on $\Om$ or $\Oms$ only.
The extension is done in the obvious way by padding with zeros
so that eq.~(3.1), for example, may be written as
\equation{
   D=\DOm+\DOms+\DdOm+\DdOms.
   \enum
}
Similarly the further decompositions
into block operators read
\equation{
   \DOm+\DOms=\sum_{{\rm all}\;\B}\DB,
   \enum
   \next{2ex}
   \DdOm=\sum_{{\rm black}\;\B}\DdB,
   \qquad
   \DdOms=\sum_{{\rm white}\;\B}\DdB,
   \enum
}
where $\DB$ denotes the Wilson--Dirac operator on the block $\B$
with Dirichlet boundary conditions and $\DdB$ the sum of the
hopping terms that move the field components
on the exterior boundary $\dB$ of the block $\B$ to its
interior boundary points.

\subsection 3.2 Quark determinant

The factorization 
\equation{
  \det D=\det\DOm\det\DOms
  \det\left\{1-\DOm^{-1}\DdOm\DOms^{-1}\DdOms\right\}
  \enum
}
is now deduced from the block structure (3.1) 
as in the case of the even--odd preconditioning
considered in subsect.~2.1.
However, contrary to what might be suspected,
the operator in the curly bracket is not quite the same as the
Schwarz-preconditioned Wilson--Dirac operator of 
ref.~[\ref{SchwarzSolver}], with cycle number $n_{\rm cy}=1$,
but their determinants coincide and 
the factorization (3.2) may therefore be regarded
as the one that is naturally associated 
to the Schwarz preconditioner.

Some further reductions and factorizations of the quark determinant
are still possible at this point. The decomposition
(3.3), for example, leads to the identity
\equation{
  \det\DOm\det\DOms=\prod_{{\rm all}\;\B}\det\DhatB,
  \enum
}
in which $\DhatB$ stands for the even--odd preconditioned Wilson--Dirac
operator on the block $\B$
with Dirichlet boundary conditions.
Another observation is that the operator in the curly bracket
in eq.~(3.5) (the Schur complement)
acts non-trivially on only those components of the quark fields that reside
on $\dOms$. Its determinant can therefore be reduced to the
space of all fields supported on this subset of points.
As explained in appendix B, a reduction to an even smaller subspace
$\VdOms$ of fields is in fact possible when
the detailed properties of
$\DdOms$ are taken into account.

Contact with the general form of the preconditioned
HMC algorithm discussed in the previous section
is now made by setting $n=2$ and 
\equation{
   R_1=\sum_{{\rm all}\;\B}\DhatB,
   \enum
   \next{2ex}
   R_2=1-\PdOms\DOm^{-1}\DdOm\DOms^{-1}\DdOms,
   \enum
}
where $\PdOms$ denotes the orthonormal projector
to the subspace $\VdOms$ (appendix B). In particular, $R_2$ is considered
to be an operator in this space,
while $R_1$ operates on quark fields that are defined on the even points of 
the full lattice.

\subsection 3.3 Block decoupling

The classical Schwarz procedure
obtains the solution of 
the Wilson--Dirac equation in an iterative process, where 
all black and all white blocks are visited alternately
and the equation is solved there
with Dirichlet boundary values given by the current approximation 
to the solution
[\ref{SchwarzSolver}].
Since the lattice Dirac operator involves only
nearest-neighbour hopping terms, the 
equations on the black blocks (and similarly those on the white blocks) 
are completely decoupled from each other and can
be solved in parallel.

In the case of the preconditioned HMC algorithm, 
a partial decoupling of
the fields on the blocks can be achieved
by restricting the molecular dynamics evolution
to a subset of all link variables, referred to as the 
{\it active link variables},
while keeping all other field variables fixed
(see fig.~2).
At the beginning or the end of every update cycle,
the gauge field should then be translated
by a random vector $v$,
\equation{
   U(x,\mu)\to U(x+v,\mu)\quad\hbox{for all}\quad x,\mu,
   \enum
}
to ensure that all link variables are treated equally on average.
Evidently,
the same can also be achieved
using several block coverings of the lattice alternately,
but this tends to be rather more complicated from the programming
point of view.

\topinsert
\vbox{
\vskip0.0cm
\epsfxsize=3.2cm\hskip4.4cm\epsfbox{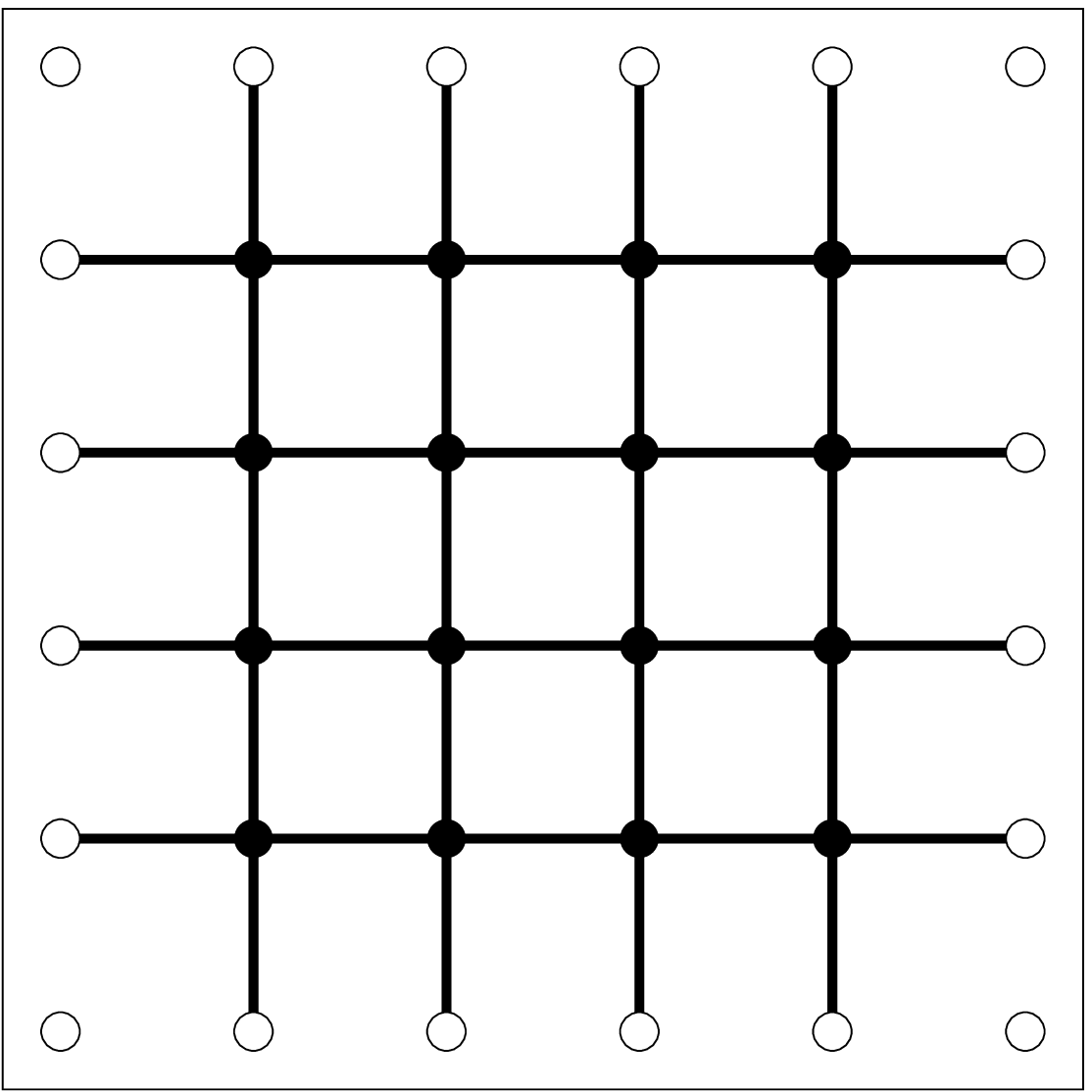}
\vskip0.4cm
\figurecaption{%
Two-dimensional view of a $6^4$ block showing the 
links on which the active link variables reside.
Such links must have
both endpoints in the same block
and at most one endpoint on the interior boundary of
the block
(open circles).
}
\vskip0.0cm
}
\endinsert

A moment of thought now reveals that
the active link variables in different blocks 
are decoupled from each other if the last term
in the hamiltonian (2.5) (the one involving the operator $R_2$)
is neglected.
The inner integrator $\Int_1(\step_2,N_0,N_1)$
therefore factorizes into 
a product of integrators, one for each block,
which evolve the active link variables
residing there as if QCD would be reduced to that block.
In particular, these parts of the integration 
can be carried out in parallel.

More important perhaps is the fact that
the restriction to the active link variables leads to 
a reduction of the
magnitude of the quark force that derives from the 
last term in the hamiltonian. This effect was anticipated
in ref.~[\ref{SchwarzAlgorithm}] and will be discussed 
again in subsect.~4.2.

\subsection 3.4 Update cycle

The generation of the next gauge-field configuration
$U'$ from the current configuration $U$ thus proceeds in the following
steps:

\vskip1ex
\noindent
(a)~First the initial momenta $\Pi(x,\mu)$ and the pseudo-fermion fields
$\phi_1(x)$ and $\phi_2(x)$
must be generated, with conditional 
probability proportional to $\rme^{-H}$. It suffices to generate
the momenta of the active link variables, since only these 
will be updated.
The field $\phi_1$ breaks up into block fields,
\equation{
   \phi_1=\sum_{{\rm all}\;\B}\phi_{\B},
   \qquad
   \phi_{\B}=\DhatB\eta_{\B},
   \enum
}
where $\eta_{\B}$ is a gaussian random field supported on the even
sites of block $\B$, and the field $\phi_2$ is similarly given
in terms of a gaussian random field $\eta_2$ through
\equation{
   \phi_2=R_2\eta_2.
   \enum
}
Both $\phi_2$ and $\eta_2$ are in the linear space $\VdOms$ of
boundary fields (appendix B).

\vskip1ex
\noindent
(b)~The molecular dynamics equations for the 
active link variables and their momenta must now be integrated
from time $0$ to time $\tau$, the trajectory length.
In the studies reported in this paper, 
the integration was performed by applying the
integrator $\Int_2(\tau,N_0,N_1,N_2)$ with some fixed
step numbers $N_0,N_1,N_2$.
The computation of the forces $F_0$ and $F_1$, which is required in this
process, is entirely standard,
while in the case of the force $F_2$ the key point to note is that
\equation{
   R_2^{-1}=1-\PdOms D^{-1}\DdOms.
   \enum
}
In particular, for all variations $\omega$ of the 
active link variables, 
\equation{
   (\omega,F_2)=
   2\kern1pt\Re\kern-2.5pt
   \left(R_2^{-1}\phi_2,\PdOms D^{-1}\delta_{\omega}DD^{-1}\DdOms\phi_2\right),
   \enum
}
and the calculation of the force thus amounts to 
solving the Wilson--Dirac equation on the full lattice for two source
fields. As already mentioned in sect.~2, any suitable preconditioner and 
solver can be used at this point. Here 
the Schwarz-preconditioned GCR solver described in 
ref.~[\ref{SchwarzSolver}]
was employed, with its own block grid chosen so as to minimize
the average execution time.

\vskip1ex
\noindent
(c)~Once the molecular dynamics equations are integrated,
the final configuration is accepted as the next
one with probability
\equation{
   \Pacc=\min\left\{1,\rme^{-\Delta H}\right\},
   \qquad \Delta H=H'-H,
   \enum
}
where $H$ and $H'$ are the values of the hamiltonian of the
initial and the final configuration respectively. 
Otherwise, i.e.~if the proposed configuration is rejected,
the old configuration is taken to be the next one.
In both cases the selected configuration is 
translated by a random vector $v$, as discussed above,
where $v$
should be chosen with some care 
(see appendix C).

\subsection 3.5 Choice of the block sizes

In principle any division of the lattice into non-overlapping 
blocks is acceptable, but the algorithm 
tends to become inefficient if the blocks are 
only a few lattice spacings wide, because in this case
only a small fraction of the link
variables get updated in each cycle.
When the lattice is divided into
$6^4$ blocks, for example, the active link 
variables are a minority of $25\%$, and this percentage
rises only slowly with the block size,
reaching values greater than $50\%$ when $12^4$ and larger blocks
are used.

Block sizes larger than $1\,\fm$ or so should also
be avoided, however, because the
Dirichlet boundary conditions then
no longer provide a safe infrared cutoff
on the spectrum of the block operators $\DhatB$.
The intended separation of the high and low modes
of the Wilson--Dirac operator
through the preconditioning is thus compromised
and the algorithm slows down
since the computation of the force $F_1$ becomes relatively
time-consuming.

To sum up, the block sizes should be 
less than $1\,\fm$, but as large as possible within this limit
so as to maximize the fraction of the active link variables.

\section 4. Tests of the algorithm

All simulations reported in this section were carried out 
on $8$ nodes ($16$ processors) of a recent PC cluster,
the same as the one described in appendix B of ref.~[\ref{SchwarzSolver}].
The general programming strategies discussed in that paper
were adopted here too and some further implementation details are given
in appendix D.

\subsection 4.1 Run parameters

Extensive simulations of two-flavour lattice QCD with 
the gauge and quark actions chosen here were previously
performed by the SESAM and the
T$\chi$L collaborations [\ref{SesamTXLauto}--\ref{SesamTxLQuark}]
and, more recently, in the framework of 
the GRAL project [\ref{GRALsize},\ref{Orth}]
using the even--odd preconditioned HMC algorithm or the plain HMC algorithm 
with an ll-SSOR preconditioner [\ref{SSOR}] for the solver.
Since most results of these studies were obtained at coupling
$\beta=6/g_0^2=5.6$, it was decided to carry out the
tests of the Schwarz-preconditioned algorithm at this value of the 
coupling.
Two lattices and three values of the hopping 
parameter $\kappa=(8+2m_0)^{-1}$ were considered,
the one corresponding to the smallest
quark mass being a new point (see table~1;
for notational convenience
the three runs at the smallest mass on the $32\times24^3$ lattice
are distinguished by an upper index on
$\kappa$).

\topinsert
\newdimen\digitwidth
\setbox0=\hbox{\rm 0}
\digitwidth=\wd0
\catcode`@=\active
\def@{\kern\digitwidth}
\tablecaption{Simulation parameters} 
\vskip-1.0ex
$$\vbox{\settabs\+&%
                  xxxxxxxx&xx&
                  xxxxxxxxxxxx&xx&
                  xxxxxxxx&xx&
                  xxxxxxxxxx&xx&
                  xxxxxxxx&x&
                  xxxxxxxx&\cr
\thicktablerule
\vskip1.0ex
                \+& \hfill Lattice\hfill
                 && \hfill Block size\hfill
                 && \hfill $\kappa$\hfill
                 && \hfill $N_0,N_1,N_2$\hfill
                 && \hfill $\Ntr$\hfill
                 && \hfill $\Pacc$\hfill
                 &\cr
\vskip1.0ex
\thintablerule
\vskip1.5ex
  \+& \hfill $32\times16^3$\hfill
  &&  \hfill $8^4$\hfill
  &&  \hskip0.5em$0.15750$\hfill 
  &&  \hfill $4,5,4$\hfill
  &&  \hfill $8000$\hfill
  &&  \hfill $0.81$\hfill
  &\cr
\vskip0.3ex
  \+& \hfill $$\hfill
  &&  \hfill $$\hfill
  &&  \hskip0.5em$0.15800$\hfill 
  &&  \hfill $4,5,5$\hfill
  &&  \hfill $9100$\hfill
  &&  \hfill $0.86$\hfill
  &\cr
\vskip0.3ex
  \+& \hfill $$\hfill
  &&  \hfill $$\hfill
  &&  \hskip0.5em$0.15825$\hfill 
  &&  \hfill $4,5,6$\hfill
  &&  \hfill $9400$\hfill
  &&  \hfill $0.90$\hfill
  &\cr
\vskip0.5ex
  \+& \hfill $32\times24^3$\hfill
  &&  \hfill $8\times6^2\times12$\hfill
  &&  \hskip0.5em$0.15750$\hfill 
  &&  \hfill $4,5,5$\hfill
  &&  \hfill $4000$\hfill
  &&  \hfill $0.82$\hfill
  &\cr
\vskip0.3ex
  \+& \hfill $$\hfill
  &&  \hfill $$\hfill
  &&  \hskip0.5em$0.15800$\hfill 
  &&  \hfill $4,5,6$\hfill
  &&  \hfill $3950$\hfill
  &&  \hfill $0.80$\hfill
  &\cr
\vskip0.3ex
  \+& \hfill $$\hfill
  &&  \hfill $$\hfill
  &&  \hskip0.5em$0.15825^{\rm a}$\hfill 
  &&  \hfill $4,5,7$\hfill
  &&  \hfill $@800$\hfill
  &&  \hfill $0.71$\hfill
  &\cr
\vskip0.3ex
  \+& \hfill $$\hfill
  &&  \hfill $$\hfill
  &&  \hskip0.5em$0.15825^{\rm b}$\hfill 
  &&  \hfill $4,5,8$\hfill
  &&  \hfill $2300$\hfill
  &&  \hfill $0.78$\hfill
  &\cr
\vskip0.3ex
  \+& \hfill $$\hfill
  &&  \hfill $$\hfill
  &&  \hskip0.5em$0.15825^{\rm c}$\hfill 
  &&  \hfill $@4,5,10$\hfill
  &&  \hfill $1100$\hfill
  &&  \hfill $0.87$\hfill
  &\cr
\vskip0.3ex
\vskip1.0ex
\thicktablerule
}
$$
\vskip-2ex
\endinsert

In physical units the lattice spacings on these lattices
decrease monotonically 
with the quark mass from $0.085\,\fm$ to about $0.078\,\fm$
[\ref{SesamTxLrzero}]
if the Sommer radius $r_0=0.5\,\fm$ [\ref{SommerScale}]
is used as the basic reference scale\kern1.5pt%
\footnote{$\dagger$}{\footnotefont%
If the mass of the $\rho$ meson was used instead, slightly
larger values would be obtained.
However, since the $\rho$ meson is a resonance and since its mass
depends quite strongly on the quark mass,
this way of setting the scale
tends to be somewhat ambiguous in full QCD and subject to potentially 
large finite-size effects.}. 
The bare current-quark masses are then
approximately $63,35$ and $21\,\MeV$ (see subsect.~4.6).
Cutoff effects can still be fairly large at these lattice spacings,
and a more meaningful statement about the 
quark masses would evidently have to include the appropriate 
renormalization factors, but these figures nevertheless 
provide a rough characterization of
the physical situation at the simulation points.

Following ref.~[\ref{SesamTXLauto}] the trajectory length $\tau$ was set
to $0.5$ in all cases. The mean distance in group space covered by the 
active link variables in the course of a trajectory
is then fairly large already (about a third of the maximal distance).
Initial runs of a few thousand trajectories were
made on all lattices to ensure a safe thermalization,
and $\Ntr$ trajectories were generated 
thereafter, as quoted in table~1.

\subsection 4.2 Magnitude of the forces $F_k$

To a first approximation it seems reasonable to fix the step numbers
$N_0,N_1,N_2$ in such a way that the increments 
$\step_0F_0,\step_1F_1,\step_2F_2$
are roughly of the same size.
The magnitude of the forces 
was determined in the test runs, and the results of these calculations
on the smaller lattices are shown in fig.~3.
For the norm of an element $X$ of the 
Lie algebra of $\SUthree$, the convention
$\|X\|^2=-2\kern1pt\tr\{X^2\}$
was adopted here, and an average was taken over all gauge configurations and
all links at the specified distance from the block boundary.

The magnitudes of the forces are thus strongly ordered,
approximately like $32\kern-1pt:\kern-1pt8\kern-1pt:\kern-1pt1$, and
$\|F_2\|$ decreases rapidly as one moves towards the centre of 
the block.
Moreover, a significant mass dependence is only seen in 
this force and only at the links where it is already very small. 
This behaviour can be understood
by noting that the expression on the 
right of eq.~(3.13) involves two quark propagators
from the boundaries $\dOms$ and $\dOm$
to the link where the force is evaluated.
The decay of the force away from the block boundary
is then essentially a consequence of the corresponding property
of the propagators.

\topinsert
\vbox{
\vskip0.0cm
\epsfxsize=7.0cm\hskip2.1cm\epsfbox{frc.eps}
\vskip0.4cm
\figurecaption{%
Average magnitude of the forces $F_k$ at the end
of the molecular dynamics trajectories on the 
$32\times16^3$ lattice at distance $d$
from the block boundary.
Dotted lines are drawn to guide the eyes, and the three curves
for $F_2$ correspond to the different quark masses
(the mass dependence of the other curves is negligible).
}
\vskip0.0cm
}
\endinsert

Figure~3 suggests to set $N_0=4,N_1=8$ and
to adjust $N_2$ so that a good acceptance rate is achieved.
Some experimenting showed, however, that $N_1=5$ 
appears to be a better choice on the lattices that were simulated.
For the values of $N_2$ quoted in table~1, the step size
$\step_2$ was then in the range $0.05-0.13$ and the average magnitude of the 
increments $\step_kF_k$ was less than $0.031$ in all cases.

\subsection 4.3 Residues \& reversibility

The computation of the quark forces $F_1$ and $F_2$ 
requires the solution of the
Wilson--Dirac equation on the blocks $\B$ and on the full lattice,
for two source fields in each case.
On the full lattice, 
the Schwarz-precon\-di\-tioned GCR solver [\ref{SchwarzSolver}]
was applied two times to the Wilson--Dirac equation directly, while
on each block $\B$ the solutions were obtained simultaneously
by solving the normal equation
\equation{
   A\psi=\eta, 
   \qquad A\equiv\DhatB(\DhatB)^{\dagger},
   \enum
}
using the conjugate-gradient (CG) algorithm.
In this case the algorithm was stopped when
the approximate solution $\psi$ satisfied
\equation{
   \|\eta-A\psi\|\leq\resF_1\|\eta\|
   \enum
}
for some specified tolerance $\resF_1$.
The analogous stopping criterion was applied,
with tolerance $\resF_2$,
when the equation on the full lattice was solved.

At the beginning and the end of the molecular dynamics trajectories,
the Wilson--Dirac equation must be solved a few more times,
both on the blocks and on the full lattice, with 
tolerances $\resS_1$ and $\resS_2$ respectively.
The BiCGstab algorithm [\ref{BiCGstabI},\ref{BiCGstabII}]
was used here to solve the block equations,
since the solution is required for one source field only
so that the application of the CG algorithm would be wasteful.

The tolerances chosen in the test runs,
\equation{
   \resF_1,\resF_2,\resS_1,\resS_2=
   \cases{10^{-7},10^{-6},10^{-10},10^{-9}  
          & on the $32\times16^3$ lattices, \cr
          \noalign{\vskip1.5ex}
          10^{-8},10^{-7},10^{-11},10^{-10} 
          & on the $32\times24^3$ lattices, \cr}
   \enum
}
are sufficiently small to guarantee the reversibility of 
the molecular dynamics trajec\-tories 
to high precision.
The maximal absolute deviation 
of the components of the 
link variables, for example, which was ever
observed after a return trajectory
was only $7\times10^{-9}$ on the smaller and $7\times10^{-10}$ on the
larger lattices, while in the case of the hamiltonian  
the differences were less than $3\times10^{-5}$ and 
$5\times10^{-6}$, respectively.

\subsection 4.4 Stability

It is well known that instabilities in
the numerical integration of the molecular dynamics equations 
may occur at small quark masses, which 
lead to violent fluctuations in $\Delta H$
and to a stagnation of the algorithm in extreme cases.
Choosing smaller integration step sizes usually cures the 
problem, but also increases the computer time required per 
trajectory
(see refs.~[\ref{UKQCDlight},\ref{CPPACSsmall}] for example).

\topinsert
\vbox{
\vskip0.0cm
\epsfxsize=10.0cm\hskip1.0cm\epsfbox{history.eps}
\vskip0.3cm
\figurecaption{%
Histories of $\Delta H$ 
and of the deviation $\Delta P$ of the average plaquette from its mean value
at $\kappa=0.15800$ on the $32\times24^3$ lattice as a function of
the trajectory number. 
Only the values 
after every 5th trajectory are plotted.
}
\vskip0.0cm
}
\endinsert

In the tests of the Schwarz-preconditioned algorithm,
severe integration instabilities were only rarely seen
in spite of the fact that the 
step size $\step_2$ was set to relatively large values.
A small algorithmic modification,
referred to as the {\it replay trick}\/ [\ref{ReplayTrick}],
was nevertheless included in order to safely avoid
periods of stagnation.
Essentially the modification amounts to 
replaying the trajectories with 
values of $|\Delta H|$ above
a specified threshold, using a smaller step size
and a somewhat complicated acceptance rule that preserves 
detailed balance. 

For illustration, a typical time series of $\Delta H$ 
(after including the replay trick) is shown in fig.~4.
The replay threshold was set to $1.5$ in this case and the overhead
in computer time that was caused by the trajectory replays 
was about $2\%$. Also shown in the figure are the fluctuations
of the average plaquette
\equation{
   P={1\over 3N_p}\sum_p\,\Re\tr\!\left\{U_p\right\},
   \enum
}
where the sum runs over all $N_p$ plaquettes
on the lattice and $U_p$ denotes the ordered product of the link
variables around the plaquette $p$.
As will be discussed shortly, these fluctuations are quite correlated,
but otherwise do not seem to have any special features.
In particular, the associated histogram has 
a nearly gaussian shape.

The fact that the (massive) Wilson--Dirac operator is not protected
from having arbitrarily small eigenvalues was always a source of concern
and may well be the cause for the integration instabilities [\ref{CPPACSsmall}].
An interesting quantity to consider in this connection is
the average number $\Nkv$ of Schwarz-preconditioned GCR iterations
that are required for the solution of the Wilson--Dirac equation
on the full lattice along a given trajectory.
While catastrophically large iteration numbers were never observed,
the distributions shown in fig.~5 have a tail 
that extends to fairly high values of $\Nkv$ at the smallest quark mass.
The replay overhead
on the $32\times24^3$ lattice was actually as large as
$20$--$30\%$ at this mass, except in the last run
where the replay threshold was set to $3.0$
and no trajectories were replayed.

\topinsert
\vbox{
\vskip0.0cm
\epsfxsize=11.0cm\hskip0.5cm\epsfbox{histo_nkv.eps}
\vskip0.2cm
\figurecaption{%
Probability distribution of the average iteration 
number $\Nkv$ per application of the
Schwarz-preconditioned GCR solver along a given trajectory.
From left to right the results obtained on the $32\times24^3$
lattice at 
$\kappa=0.15750$, $0.15800$, $0.15825^{\rm b}$
are shown.
}
\vskip-0.0cm
}
\endinsert

\subsection 4.5 Autocorrelation times

An estimation of the relevant auto\-correlation times
is clearly essential in the present context, since
the efficiency of the algorithm can only be determined when
these are known. Unfortunately the available statistics is 
insufficient for a solid ``error of the error" analysis
[\ref{UlliError}],
and the values for the integrated auto\-correlation times 
quoted below should therefore be taken as first estimates only.

The auto\-correlation functions 
of the average plaquette $P$ and the average number $\Nkv$
of GCR iterations are plotted in fig.~6.
An approximation to the four-point auto\-correlation function
introduced by Madras and Sokal [\ref{MadrasSokal}] 
was used here to determine the statistical errors
(see appendix E).
As usual the associated integrated auto\-correlation times 
are obtained by summing the auto\-correlation functions
up to some maximal time lag, referred to as
the summation window.
The particular prescription adopted here was to 
truncate the sums at 
the first point where the auto\-correlation
function vanishes within errors.
In general this rule yields consistent results and
summation windows a few times larger than the calculated
autocorrelation times.

\topinsert
\vbox{
\vskip0.0cm
\epsfxsize=11.4cm\hskip0.3cm\epsfbox{auto.eps}
\vskip0.4cm
\figurecaption{%
Normalized autocorrelation functions of the average plaquette $P$ 
(dark error band) and the average number $\Nkv$ of GCR iterations
(grey band). In the first and the second row of plots
the lattice size is $32\times16^3$ 
and $32\times24^3$ respectively,
and the hopping parameters increase from left to right as in fig.~5.
}
\vskip-0.2cm
}
\endinsert

The values of the integrated auto\-correlation times listed in table~2
seem to be quite a bit higher than those obtained
in previous studies of two-flavour QCD, but 
the comparison is not straightforward and must in any case 
remain uncertain to some extent in view of 
the large statistical errors (see subsect.~4.7).
It should also be noted that the solver iteration numbers
are usually a worst case, and more physical quantities (hadron masses
in particular) typically have much smaller integrated
auto\-correlation times. 

Whether there is a systematic trend in the auto\-correlation times
as a function of the quark mass or the lattice size is difficult 
to tell from the figures listed in table~2.
There is, however, no indication that the auto\-correlation
times grow when the quark mass is lowered.
Some of the runs were perhaps a bit too short
to safely exclude biased results,
and longer simulations
may be required to firmly establish
the decrease of the auto\-correlation times 
as a function of the step size $\step_2$ suggested by the last
three entries in the table.

\topinsert
\newdimen\digitwidth
\setbox0=\hbox{\rm 0}
\digitwidth=\wd0
\catcode`@=\active
\def@{\kern\digitwidth}
\tablecaption{Integrated autocorrelation times$\,^{\ast}$} 
\vskip1.0ex
$$\vbox{\settabs\+&%
                  xxxxxxxx&xx&
                  xxxxxxxx&xx&
                  xxxxxxxxxxxxxx&xx&
                  xxxxxxxxxxxxxx&\cr
\thicktablerule
\vskip1.0ex
                \+& \hfill Lattice\hfill
                 && \hfill $\kappa$\hfill
                 && \hfill $\tauint[P]$\hfill
                 && \hfill $\tauint[\Nkv]$\hfill
                 &\cr
\vskip1.0ex
\thintablerule
\vskip1.5ex
  \+& \hfill $32\times16^3$\hfill
  &&  \hskip0.5em$0.15750$\hfill 
  &&  \hfill $68(25)$\hfill
  &&  \hfill $168(42)$\hfill
  &\cr
\vskip0.3ex
  \+& \hfill $$\hfill
  &&  \hskip0.5em$0.15800$\hfill 
  &&  \hfill $32(7)@$\hfill
  &&  \hfill $162(56)$\hfill
  &\cr
\vskip0.3ex
  \+& \hfill $$\hfill
  &&  \hskip0.5em$0.15825$\hfill 
  &&  \hfill $57(18)$\hfill
  &&  \hfill $135(39)$\hfill
  &\cr
\vskip0.5ex
  \+& \hfill $32\times24^3$\hfill
  &&  \hskip0.5em$0.15750$\hfill 
  &&  \hfill $53(22)$\hfill
  &&  \hfill $144(51)$\hfill
  &\cr
\vskip0.3ex
  \+& \hfill $$\hfill
  &&  \hskip0.5em$0.15800$\hfill 
  &&  \hfill $33(11)$\hfill
  &&  \hfill $122(36)$\hfill
  &\cr
\vskip0.3ex
  \+& \hfill $$\hfill
  &&  \hskip0.5em$0.15825^{\rm a}$\hfill 
  &&  \hfill $33(13)$\hfill
  &&  \hfill $84(24)@$\hfill
  &\cr
\vskip0.3ex
  \+& \hfill $$\hfill
  &&  \hskip0.5em$0.15825^{\rm b}$\hfill 
  &&  \hfill $19(5)@$\hfill
  &&  \hfill $65(20)@$\hfill
  &\cr
\vskip0.3ex
  \+& \hfill $$\hfill
  &&  \hskip0.5em$0.15825^{\rm c}$\hfill 
  &&  \hfill $12(4)@$\hfill
  &&  \hfill $22(6)@@$\hfill
  &\cr
\vskip0.3ex
\vskip1.0ex
\thicktablerule
\vskip1.0ex
\+{\footnotefont$^{\ast}$ In numbers of trajectories of length $\tau=0.5$}&\cr
}
$$
\vskip0.5ex
\endinsert

\subsection 4.6 Quark and pion masses

In this algorithmic study,
the computation of the current-quark mass and the pion mass
is not of central interest,
but it helps to determine the physical situation on the 
simulated lattices, as discussed at the beginning of this section.

Since a very accurate determination of the masses is not
required, error reduction techniques were not applied and 
the masses were obtained straightforwardly from the 
two-point functions of the isospin axial current and density.
In terms of the quark field $\psi$ and 
the antiquark field $\psibar$, the latter are given by
\equation{
   A_{\mu}^a=\psibar\dirac{\mu}\dirac{5}\frac{1}{2}\tau^a\psi,
   \qquad
   P^a=\psibar\dirac{5}\frac{1}{2}\tau^a\psi,
   \enum
}
where $\tau^1,\tau^2,\tau^3$ denote the isospin Pauli matrices.
As usual the two-point functions were evaluated
at vanishing spatial momentum,
\equation{
   \sum_{x_1,x_2,x_3}\left\langle P^a(x)P^b(0)\right\rangle\equiv
   \frac{1}{2}\delta^{ab}\fpp(x_0),
   \enum
   \next{1ex}
   \sum_{x_1,x_2,x_3}\left\langle A^a_0(x)P^b(0)\right\rangle\equiv
   \frac{1}{2}\delta^{ab}\fap(x_0),
   \enum
}
and the appropriate signed average of the calculated functions
at time $x_0$ and $T-x_0$ was taken.

In the test runs, the gauge-field configurations
generated after every $100$ update cycles were saved to disk,
allowing physical quantities such as the quark and the pion mass to be
calculated later.
The fields in these ensembles are only weakly
correlated, and no systematic
increase in the naive statistical errors was in fact
observed when the measured values of the two-point functions 
$\fpp(x_0)$ and $\fap(x_0)$  
were averaged over small bins of successive configurations.
For the mass calculations the values of the two-point functions
were therefore assumed to be statistically independent and 
the results were checked by repeating the calculations with
the data averaged over bins of $2$ or $3$ measurements.

The conservation of the axial current in the continuum limit
implies that the ratio
\equation{
  \{\fap(x_0+1)-\fap(x_0-1)\}\{4\fpp(x_0)\}^{-1}
  \enum
}
is independent of $x_0$ and equal to the bare
current-quark mass $m$, up to terms that vanish proportionally
to the lattice spacing. 
In the range $8\leq x_0\leq16$ the data for the ratio
were actually found to be constant within errors. 
The quark mass was then extracted through a correlated least-squares
fit of the ratio in this range,
using jackknife error estimates for the covariance matrix.

\topinsert
\newdimen\digitwidth
\setbox0=\hbox{\rm 0}
\digitwidth=\wd0
\catcode`@=\active
\def@{\kern\digitwidth}
\tablecaption{Average plaquette, current-quark mass and pion mass} 
\vskip-1.0ex
$$\vbox{\settabs\+&%
                  xxxxxxxx&xx&
                  xxxxxxxxxx&xx&
                  xxxxxxxxxxxxx&xx&
                  xxxxxxxxxxx&x&
                  xxxxxxxxxxx&\cr
\thicktablerule
\vskip1.0ex
                \+& \hfill Lattice\hfill
                 && \hfill $\kappa$\hfill
                 && \hfill $P$\hfill
                 && \hfill $am$\hfill
                 && \hfill $a\mpi$\hfill
                 &\cr
\vskip1.0ex
\thintablerule
\vskip1.5ex
  \+& \hfill $32\times16^3$\hfill
  &&  \hfill $0.15750$\hfill 
  &&  \hfill $0.57255(6)$\hfill
  &&  \hfill $0.0279(4)$\hfill
  &&  \hfill $0.298(5)$\hfill
  &\cr
\vskip0.3ex
  \+& \hfill $$\hfill
  &&  \hfill $0.15800$\hfill 
  &&  \hfill $0.57335(4)$\hfill
  &&  \hfill $0.0153(4)$\hfill
  &&  \hfill $0.242(4)$\hfill
  &\cr
\vskip0.3ex
  \+& \hfill $$\hfill
  &&  \hfill $0.15825$\hfill 
  &&  \hfill $0.57387(5)$\hfill
  &&  \hfill $0.0092(4)$\hfill
  &&  \hfill $0.209(7)$\hfill
  &\cr
\vskip0.5ex
  \+& \hfill $32\times24^3$\hfill
  &&  \hfill $0.15750$\hfill 
  &&  \hfill $0.57250(4)$\hfill
  &&  \hfill $0.0273(4)$\hfill
  &&  \hfill $0.280(3)$\hfill
  &\cr
\vskip0.3ex
  \+& \hfill $$\hfill
  &&  \hfill $0.15800$\hfill 
  &&  \hfill $0.57344(3)$\hfill
  &&  \hfill $0.0143(3)$\hfill
  &&  \hfill $0.188(5)$\hfill
  &\cr
\vskip0.3ex
  \+& \hfill $$\hfill
  &&  \hskip1.0em$0.15825^{\rm all}$\hfill 
  &&  \hfill $0.57383(2)$\hfill
  &&  \hfill $0.0084(4)$\hfill
  &&  \hfill $0.153(4)$\hfill
  &\cr
\vskip0.3ex
\vskip1.0ex
\thicktablerule
}
$$
\vskip-2ex
\endinsert

The pion mass $\mpi$ was calculated in the same way
by fitting the effective mass $m_{\rm eff}(x_0)$
at large times $x_0$.
At any given $x_0$, the effective mass is obtained by solving the 
equation
\equation{
  {h(x_0-1)\over h(x_0)}
  ={\fpp(x_0-1)\over\fpp(x_0)},
  \qquad h(x_0)\equiv\rme^{-x_0M}+\rme^{-(T-x_0)M},
  \enum
}
for $M$ and setting $m_{\rm eff}(x_0)=M$. The associated 
covariance matrix was again cal\-cu\-lated using jackknife bins and
the fit ranges were chosen so that a good fit quality was obtained.

Where a comparison is possible,
the results listed in table~3 coincide with those published
by the SESAM, T$\chi$L and GRAL collaborations
[\ref{SesamTXLauto}--\ref{Orth}]
within $1$ or $2$ sigma, an exception being a
deviation by $3$ sigma of the pion mass at $\kappa=0.15750$ on the 
$32\times16^3$ lattice.
The quark masses change only slightly with the
lattice size, as it should be close to the continuum limit,
while the size-dependence of the pion mass is not
small on these lattices [\ref{GRALsize},\ref{Orth}].
In physical units the quark and the pion masses quoted on the last line of
table~3 are about $21\,\MeV$ and $386\,\MeV$ respectively 
(if the scale is set through the Sommer radius).

\subsection 4.7 Speed and efficiency of the algorithm

In practice the average number of accepted 
gauge-field configurations generated
per day on the computer that is used for the simulations
is a relevant performance figure.
As can be seen from fig.~7, the experience made 
with the Schwarz-preconditioned HMC is quite encouraging in this respect,
given the fact that the tests were performed on a relatively 
small machine.
In the most time-critical parts of the program,
the machine delivered about $26$ Gflop/s [\ref{SchwarzSolver}],
and the total computer time spent for the tests was a little more than
half a year.

\topinsert
\vbox{
\vskip0.0cm
\epsfxsize=6.8cm\hskip2.6cm\epsfbox{tim.eps}
\vskip0.4cm
\figurecaption{%
Average number $\langle N_{\rm cf}\rangle$ of accepted 
gauge-field configurations generated 
per day by the Schwarz-preconditioned HMC algorithm
on $8$ nodes of a recent PC cluster
(described in ref.~[\ref{SchwarzSolver}], appendix B),
as a function of the bare current-quark mass.
The point at the smallest mass on the $32\times24^3$ lattice is from
the last run in table~1
and the linear scaling curves (dotted lines) 
are drawn for comparison with the data.
}
\vskip0.0cm
}
\endinsert

At the smallest quark mass that was considered, most of the time is used
for the solution of the Wilson--Dirac equation on the full lattice.
The situation is different at the larger quark masses,
where the solver converges faster while the time spent in
the other parts of the program is nearly unchanged.
For this reason, and since the step number $N_2$ had to be adjusted slightly,
the configuration production rate does not follow a simple scaling law
with respect to the quark mass. In particular, the three lower points
in fig.~7 are only accidentally fitted by the dotted line.

An interesting measure for the efficiency of the Schwarz-preconditioned 
algorithm is provided by the cost figure 
\equation{
  \nu=10^{-3}\left(2N_2+3\right)\tint[P].
  \enum
}
On average, this is the number 
of times the Wilson--Dirac equation must be solved on the full lattice,
in units of thousands,
before a statistically independent value of the average plaquette 
$P$ is obtained.
Evidently $\nu$ is independent of the machine, the program
and the particular solver which are used.
The numbers listed in table~4 show that   
the algorithm is remarkably stable in these terms, and
there certainly is no indication, in the parameter range considered
and within the errors quoted, 
that it would
slow down towards smaller quark masses or larger lattices.

\topinsert
\newdimen\digitwidth
\setbox0=\hbox{\rm 0}
\digitwidth=\wd0
\catcode`@=\active
\def@{\kern\digitwidth}
\tablecaption{Values of the cost figure $\nu$} 
\vskip1.3ex
$$\vbox{\settabs\+&%
                  xxxxxxxx&xx&
                  xxxxxxxxxx&xx&
                  xxxxxxxxxx&\cr
\thicktablerule
\vskip1.0ex
                \+& \hfill Lattice\hfill
                 && \hfill $\kappa$\hfill
                 && \hfill $\nu$\hfill
                 &\cr
\vskip1.0ex
\thintablerule
\vskip1.5ex
  \+& \hfill $32\times16^3$\hfill
  &&  \hfill $0.15750$\hfill 
  &&  \hskip0.8em$0.75(28)$\hfill
  &\cr
\vskip0.3ex
  \+& \hfill $$\hfill
  &&  \hfill $0.15800$\hfill 
  &&  \hskip0.8em$0.42(9)$\hfill
  &\cr
\vskip0.3ex
  \+& \hfill $$\hfill
  &&  \hfill $0.15825$\hfill 
  &&  \hskip0.8em$0.86(27)$\hfill
  &\cr
\vskip0.5ex
  \+& \hfill $32\times24^3$\hfill
  &&  \hfill $0.15750$\hfill 
  &&  \hskip0.8em$0.69(29)$\hfill
  &\cr
\vskip0.3ex
  \+& \hfill $$\hfill
  &&  \hfill $0.15800$\hfill 
  &&  \hskip0.8em$0.50(17)$\hfill
  &\cr
\vskip0.3ex
  \+& \hfill $$\hfill
  &&  \hskip1.0em$0.15825^{\rm a}$\hfill 
  &&  \hskip0.8em$0.56(22)$\hfill
  &\cr
\vskip0.3ex
  \+& \hfill $$\hfill
  &&  \hskip1.0em$0.15825^{\rm b}$\hfill 
  &&  \hskip0.8em$0.36(10)$\hfill
  &\cr
\vskip0.3ex
  \+& \hfill $$\hfill
  &&  \hskip1.0em$0.15825^{\rm c}$\hfill 
  &&  \hskip0.8em$0.28(9)$\hfill
  &\cr
\vskip1.0ex
\thicktablerule
}
$$
\vskip-0.0ex
\endinsert

If another or no preconditioner is used for the
HMC algorithm, the cost figure $\nu$ may be defined 
in the same way, with $N_2$ replaced by the appropriate step number.
In refs.~[\ref{SesamTXLauto}--\ref{Orth}], for example,
the plain HMC algorithm was studied at the same gauge coupling
and hopping parameters $\kappa=0.15750$ and $\kappa=0.15800$ 
which were considered here.
On the $32\times16^3$ lattice
the corresponding values of $\nu$, 
as calculated from 
the run parameters and auto\-correlation times quoted in ref.~[\ref{Orth}],
are equal to $1.0(6)$ and $2.8(8)$ respectively.
These numbers rise to $1.8(8)$ and $5.1(5)$ 
on the $40\times24^3$ lattice, which shows quite clearly
that the efficiency of the plain HMC algorithm decreases
when the lattice size is increased or if the quark mass is taken
to smaller values.
In particular,
already at $\kappa=0.15800$ (where the pion mass is approximately $465\,\MeV$)
the Schwarz-preconditioned algorithm is much more efficient.

Although different lattice formulations were chosen in most cases, it
is interesting to note that large values of $\nu$ are common to all
previous simulations of two-flavour QCD at small quark masses. In a
study conducted by the UKQCD collaboration [\ref{UKQCDlight}], for
example, using the even--odd preconditioned HMC algorithm, the cost
figure on a $32\times16^3$ lattice at $\mpi\simeq420\,\MeV$ was $5.5(11)$.
Significantly smaller masses were reached by the
CP-PACS collaboration on a coarse $24\times12^3$ lattice
[\ref{CPPACSsmall}], and in these simulations $\nu$ ranged from
$9.4(18)$ at $\mpi\simeq398\,\MeV$ to $29(10)$ at $\mpi\simeq255\,\MeV$.

\section 5. Concluding remarks

The Schwarz-preconditioned HMC algorithm is intended
for simulations of lattice QCD at small lattice spacings and small
quark masses.
It is particularly well suited for parallel processing,
which can be a decisive advantage
when the continuum limit is approached
and very large lattices 
have to be simulated. Many of the theoretically expected properties
of the algorithm were confirmed in the tests reported in this paper,
and although important uncertainties remain, there are 
clear indications that 
the algorithm does indeed have a favourable scaling behaviour with 
respect to the quark mass.

So far only the standard Wilson formulation of lattice QCD
was considered, but $\rmO(a)$ improvement and 
double-plaquette terms 
can be easily included in the Schwarz-preconditioned algorithm.
Whether its excellent scaling properties will be preserved
is difficult to foresee, however, because
the autocorrelation times are unpredictable.
Independently of the lattice formulation, 
the algorithm is not expected to perform particularly well
at large quark masses or
at lattice spacings larger than $0.1\,\fm$ or so.
The problem on coarse lattices simply is that 
all choices of the block sizes tend to be inappropriate
from one or the other point of view (cf.~subsect.~3.5).

The particular Schwarz preconditioner that was considered here
is only one example of a domain decomposition preconditioner.
Close to the continuum limit
more complicated preconditioners,
based on a hierarchy of block grids perhaps or
alternative domain decompositions,
may conceivably be more efficient.
The potential for further accelerations 
along these lines is limited, however, until significantly larger
step numbers are required than those quoted in table~1.

\vskip1ex
I am indebted to Rainer Sommer for many helpful discussions
on QCD simulation algorithms
and to Ulli Wolff for correspondence on
the analysis of simulation time series.
Thanks also go to Boris Orth for sending a copy of his PhD 
thesis prior to publication, and to him and Thomas Lippert for
correspondence on
the SESAM, T$\chi$L and GRAL projects.
The simulations reported in this paper
were carried out on a PC cluster at
the Institut f\"ur Theo\-retische Physik der Universit\"at Bern,
which was funded in part by
the Schweizerischer Nationalfonds.
I thank both institutions for the continuous support
given to this project.

\appendix A. $\SUthree$ exponential function

In the course of the numerical integration of the molecular dynamics
equations, the link variables are updated by left-multiplication with 
a matrix function $\Exp(X)$, where $X$ is proportional to the 
field momentum at the given link. $\Exp(X)$ is usually taken to be
the exponential function in $\SUthree$, but there are alternative
choices that are theoretically equally acceptable and
easier to implement.

To ensure that the leap-frog integration
will be well-defined, reversible and quadra\-tically convergent,
the following properties must hold:

\vskip1ex\noindent
(1)~$\Exp(X)$ is a smooth mapping from 
the Lie algebra of $\SUthree$ to the group.

\vskip1ex\noindent
(2)~The function satisfies $\Exp(X)\Exp(-X)=1$ for all $X$.

\vskip1ex\noindent
(3)~At small $t$ the asymptotic expansion 
$\Exp(tX)=1+tX+\frac{1}{2}t^2X^2+\rmO(t^3)$ holds.

\vskip1ex\noindent
All these requirements can be met by setting
\equation{
  \Exp(X)=U_1U_2U_3U_2U_1,
  \enum
}
where $U_1$, $U_2$ and $U_3$ are certain
$\SUtwo$ rotations in the (12), (13) and
(23) subspaces respectively. 
They are obtained by representing
the anti-hermitian traceless matrix $X$ as a sum
of the three matrices
\equation{
  Y_1=\pmatrix{x_1 & X_{12} & 0 \cr
               X_{21} & -x_1 & 0 \cr
               0 & 0 & 0 \cr},
  \qquad x_1=\frac{1}{3}\left(X_{11}-X_{22}\right),
  \enum
  \next{2.0ex}
  Y_2=\pmatrix{
               x_2 & 0& X_{13}  \cr
               0 & 0 & 0 \cr
               X_{31} & 0 & -x_2 \cr
               },
  \qquad x_2=\frac{1}{3}\left(X_{11}-X_{33}\right),
  \enum
  \next{2.0ex}
  Y_3=\pmatrix{
               0 & 0 & 0 \cr
               0 & x_3 & X_{23}  \cr
               0 & X_{32} &  -x_3 \cr
               },
  \qquad x_3=\frac{1}{3}\left(X_{22}-X_{33}\right).
  \enum
}
Evidently, these are generators of the $\SUtwo$ subgroups just mentioned,
and it is then straightforward to prove that 
\equation{
  U_1=\left(1+\frac{1}{4}Y_1\right)\left(1-\frac{1}{4}Y_1\right)^{-1},
  \enum
  \next{2.0ex}
  U_2=\left(1+\frac{1}{4}Y_2\right)\left(1-\frac{1}{4}Y_2\right)^{-1},
  \enum
  \next{2.0ex}
  U_3=\left(1+\frac{1}{2}Y_3\right)\left(1-\frac{1}{2}Y_3\right)^{-1}
  \enum
}
is a possible choice of the matrices $U_k$.
The inversions in these equations can be worked out analytically
and are numerically safe, since the $Y_k$'s have purely imaginary eigenvalues.
In particular, it is straightforward to write a fast program that computes
$\Exp(X)$ for any given $X$ practically to machine precision.

\appendix B. Boundary fields

In this appendix the
space $\VdOms$ of boundary fields
is specified explicitly.
The choice of this space is actually not unique,
but it should be such that the associated orthonormal 
projector $\PdOms$ satisfies
\equation{
   \DdOms\PdOms=\DdOms.
   \enum
}
This property guarantees 
that the determinant of the Schur complement in eq.~(3.5)
coincides with the determinant of the projected operator $R_2$,
which is all what is assumed in sects.~3 and 4.

The action of the operator $\DdOms$ on an arbitrary quark field $\psi(x)$
is given by
\equation{
   \DdOms\psi(x)=-\tOms(x)\sum_{\mu=0}^3\bigl\{
   \frac{1}{2}\left(1-\dirac{\mu}\right)\tOm(x+\hat{\mu})
   U(x,\mu)\psi(x+\hat{\mu})
   \noenum
   \next{0.5ex}
   \hskip10.82em
   +\frac{1}{2}\left(1+\dirac{\mu}\right)\tOm(x-\hat{\mu})
   U(x-\hat{\mu},\mu)^{-1}\psi(x-\hat{\mu})\bigr\},
   \enum
}
where $\tOm(x)$ and $\tOms(x)$ are the characteristic functions
of $\Om$ and $\Oms$.
As illustrated by fig.~8, the 
terms on the right-hand side of 
this equation move the Dirac spinors from $\dOms$ to $\dOm$
and multiply them with the projectors $\frac{1}{2}(1\pm\dirac{\mu})$ 
and the appropriate link variables. 
In particular, two components of the spinors residing
on the subset 
\equation{
  \dOmsring=\left\{x\in\dOms\bigm| 
  \hbox{$x+\hat{\mu}\in\dOm$ or
  $x-\hat{\mu}\in\dOm$ for one value of $\mu$ only}\right\}
  \enum
}
are lost irretrievably through the application of the projectors.

\topinsert
\vbox{
\vskip0.0cm
\epsfxsize=4.0cm\hskip4.0cm\epsfbox{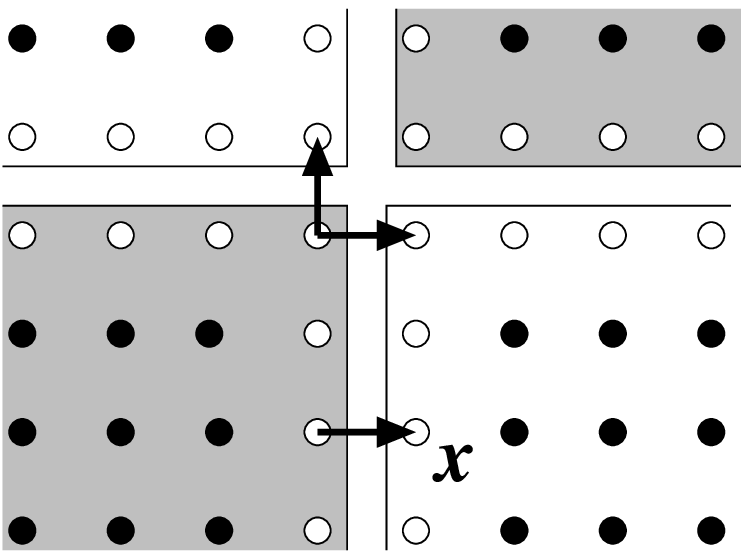}
\vskip0.3cm
\figurecaption{%
The hopping terms in $\DdOms$ move the quark spinors from the 
interior to the exterior boundary points of the black blocks.
Depending on the position of the inner point, the spinor is
moved in one or more directions.
}
\vskip0.0cm
}
\endinsert

It is now straightforward to show that 
the operator
\equation{
  \PdOms\psi(x)=\cases{
                0 
                &if $x\notin\dOms$,\cr
                \noalign{\vskip1.0ex}
                \frac{1}{2}(1+\dirac{\mu})\psi(x)
                &if $x\in\dOmsring$ and $x+\hat{\mu}\in\dOm$,\cr
                \noalign{\vskip1.0ex}
                \frac{1}{2}(1-\dirac{\mu})\psi(x)
                &if $x\in\dOmsring$ and $x-\hat{\mu}\in\dOm$,\cr
                \noalign{\vskip1.0ex}
                \psi(x)
                &otherwise,\cr
                }
  \enum
}
satisfies eq.~(B.1). Moreover, this choice excludes the trivial
nullspace of $\DdOms$ and thus minimizes the dimension of the space
$\VdOms$ of boundary quark fields.

\appendix C. Translation vectors

The field translations at the end of the update cycles
of the Schwarz-preconditioned algorithm should ideally be such that 
all link variables are visited at approximately the same rate.
In particular, long gaps between subsequent updates of 
a given link variable should be avoided as far as possible.
One might think that a purely random choice 
of the translation vectors $v$ will be acceptable from this point
of view, but the visiting frequency on some links 
can actually be
far below average in this case,  even
after hundreds of update cycles.

\topinsert
\vbox{
\vskip0.0cm
\epsfxsize=5.0cm\hskip3.5cm\epsfbox{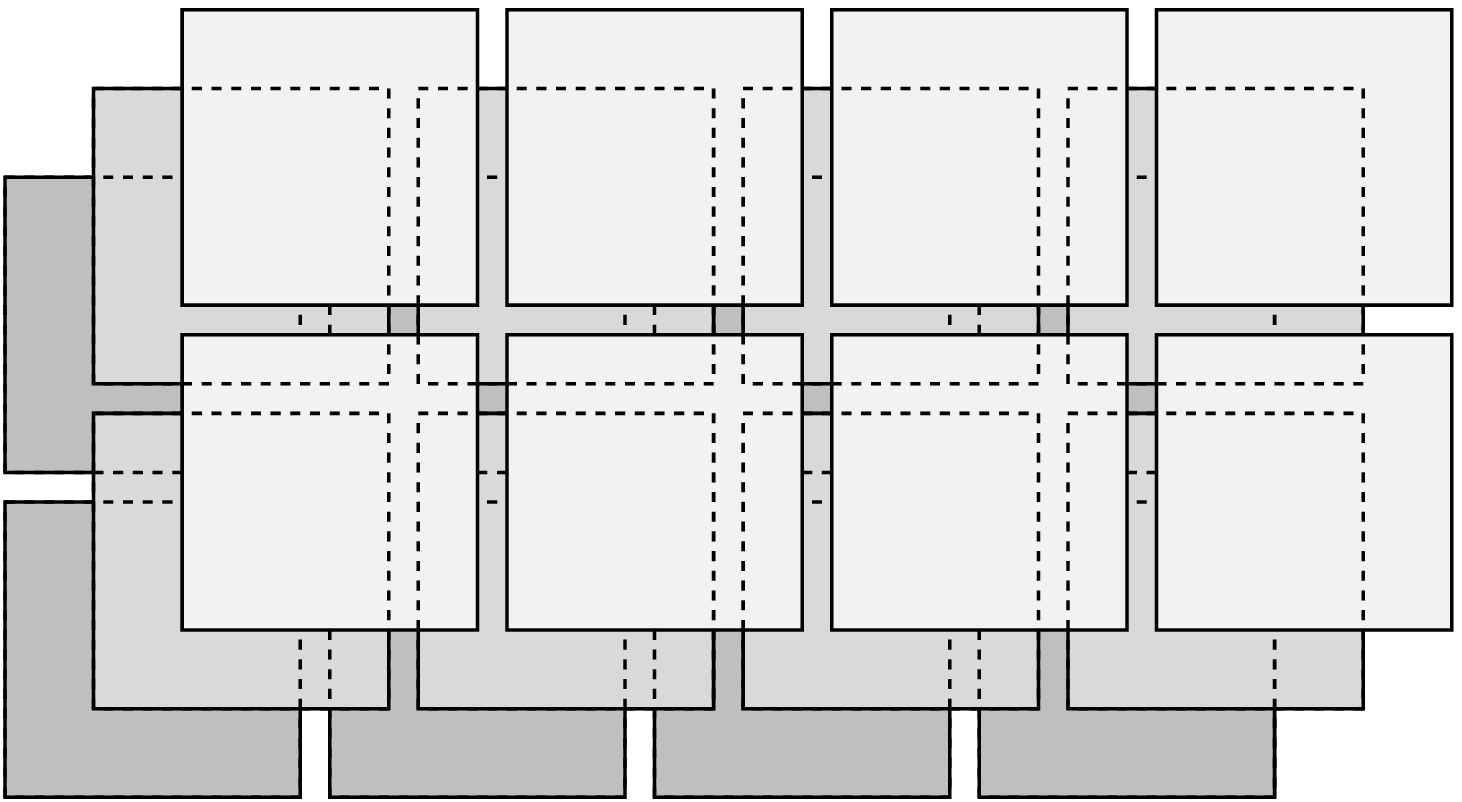}
\vskip0.3cm
\figurecaption{%
Three copies of a block grid in two dimensions, two of them
being translated along a diagonal direction at regular distances. 
Taking the boundary
conditions into account, the lattice is completely covered 
by the blocks in these grids.
}
\vskip0.0cm
}
\endinsert

More balanced visiting frequencies can be obtained
by adopting a quasi-random scheme, where a purely random choice of $v$
is followed, in the next few update cycles,
by translations along a diagonal lattice direction
in regular steps (see fig.~9). 
If the lattice is divided into blocks of size $8^4$,
for example,
performing a random translation and then three times
a translation by the vector
\equation{
   v=(2,2,2,2)
   \enum
}
yields a fairly narrow visiting distribution.
For general block sizes $b_{\mu}$, it suffices to 
take $n-1$ regular steps by the vector
\equation{
   v=(b_0,b_1,b_2,b_3)/n
   \enum
}
(or close-by integer vectors),
where $n=\frac{1}{2}\min\{10,b_0,b_1,b_2,b_3\}$.

\appendix D. Programming issues

The programming of the Schwarz-preconditioned algorithm
is not completely trivial, particularly so
on parallel computers, where the block decomposition of the lattice
gives rise to additional complications.
Object orientation and generic programming are 
useful general strategies that may be applied here [\ref{SchwarzSolver}],
and some more specific issues will now be addressed.

\subsection D.1 Parallelization

Clearly the communication overhead is minimized
if each block in the domains $\Om$ and $\Oms$
is fully contained in one of the sublattices on which
the processors of a parallel computer operate.
This was the case in the test runs reported here, but 
more flexible assignments may eventually be needed,
where the processor sublattices divide the blocks in some way. 

On machines with symmetric multiprocessing nodes, for example,
the natural choice is to contain the blocks
in the node sublattices. In this case the processors can
operate on the block fields in a shared-memory mode
and the network is only used when the global fields get involved. 
In general the communication overhead can always be reduced
by dividing the blocks into smaller blocks
that are fully contained in the processor sublattices.
The Schwarz-preconditioned GCR solver may then be set up on this
block grid and be used 
for the solution of the Wilson--Dirac equation on both the
full lattice and the big blocks.

\subsection D.2 Block fields

In the program a block may be defined through a data structure
that contains the relevant geometrical information
(embedding and nearest-neighbour indices), the local gauge field
and a set of quark fields. 
An obvious advantage of this approach is that
the operations on these fields
can proceed locally without having to 
gather and scatter any data from and to the global fields.
With a proper layout of the local fields in memory, 
streaming data accesses, and hence a higher processing speed,
are thus made possible.

Most computations are actually local operations,
an exception being the calculation of the force $F_2$, which
requires the solution of the Wilson--Dirac equation on the full lattice.
The current values of the gauge fields
on the blocks may need to be copied to the global fields at this point,
but this is a minor complication, since the fields must be copied
only once per force calculation.

\subsection D.3 Single-precision acceleration

In the course of the numerical integration of 
the molecular dynamics equations,
an exponential amplification of rounding errors is 
possible and important significance
losses are in any case unavoidable when the difference $\Delta H$ is 
calculated at the end of the trajectories.
The algorithm was therefore implemented using double-precision
(64 bit) data and arithmetic for all fields.

Through the intermediate use of single-precision data and arithmetic,
the CG and the Schwarz-preconditioned GCR solvers for the 
Wilson--Dirac equation can, however, be accelerated
without compromising the precision of the final results
[\ref{NumMethods},\ref{SchwarzSolver}].
On current PC processors, where
the relevant single-precision programs
are significantly faster than their double-precision versions,
an acceleration by a factor $2$ or so can be achieved in this way.

\appendix E. Calculation of autocorrelation times

In practice the calculation of autocorrelation times 
tends to be somewhat ambiguous unless the available time series 
are very long.
The particular prescriptions that
were used in this paper are described in the following for the 
case of a primary observable $A$ such as the average plaquette.
Essentially the discussion follows refs.~[\ref{UlliError},\ref{MadrasSokal}]
where further details and alternative strategies can be found.

\subsection E.1 Preliminaries

Let $a_1,a_2,\ldots,a_N$ be a time series of measurements of $A$
obtained in the course of a numerical simulation.
It is helpful to imagine that infinitely many such simulations
were made, with independent random numbers so that
the different runs may be assumed to be uncorrelated.
The true expectation value of $A$ is then
\equation{
  a=\langle a_i\rangle,
  \enum
}
where $\langle\ldots\rangle$ denotes the average over the infinite set
of uncorrelated simulations. This value should be distinguished
from the average
\equation{
  \abar={1\over N}\sum_{i=1}^Na_i,
  \enum
}
which is obtained in a particular simulation. From the point of view
of the set of simulations, $\abar$ is a stochastic variable with
expectation value $\langle\abar\rangle=a$.

In terms of the true autocorrelation function
\equation{
   \Gamma(t)=\Gamma(-t)=
   \left\langle(a_i-a)(a_{i+t}-a)\right\rangle,
   \enum
}
the statistical variance of the measured value $\abar$ of $A$ is
given by
\equation{
   \sigma^2=\left\langle(\abar-a)^2\right\rangle={1\over N^2}
   \sum_{i,j=1}^N\Gamma(i-j).
   \enum
}
At large $N$ this equation may be written in the form
\equation{
   \sigma^2=2\tint\sigma_0^2
   \enum
}
where $\sigma_0^2=\Gamma(0)/N$ is the naive variance and
\equation{
  \tint={1\over2}+\sum_{t=1}^{\infty}{\Gamma(t)\over\Gamma(0)}
  \enum
}
the integrated autocorrelation time.

\subsection E.2 Madras--Sokal approximation

The expression
\equation{
   \Gbar(t)={1\over N-t}\sum_{i=1}^{N-t}
   (a_i-\abar)(a_{i+t}-\abar)
   \enum
}
provides an approximation to the true autocorrelation function,
which can be computed from the available time series.
Its expectation value 
coincides with $\Gamma(t)$ up to terms that are negligible (at large $N$)
with respect to the deviation
\equation{
  \delta\Gamma(t)=\Gbar(t)-\left\langle\Gbar(t)\right\rangle,
  \enum
}
whose variance determines the
statistical error of $\Gbar(t)$.

The covariance of $\delta\Gamma(t)$ involves 
a sum over the four-point autocorrelation function, 
which is usually difficult to compute
precisely from the available data alone.
However,
as first noted by Madras and Sokal [\ref{MadrasSokal}],
the disconnected parts of the four-point function make the dominant 
contribution to the sum, and
if only these are kept, the covariance becomes
\equation{
  \left\langle\delta\Gamma(t)\delta\Gamma(s)\right\rangle\simeq
  {1\over N}\sum_{k=-\infty}^{\infty}
  \bigl\{\Gamma(k)\Gamma(k+t-s)+
  \Gamma(k+t)\Gamma(k-s)\bigr\},
  \enum
}
where $t,s\ll N$ was assumed and any subleading terms were neglected.
It is then also possible to derive the elegant formula
\equation{
  \left\langle\delta\rho(t)^2\right\rangle\simeq
  {1\over N}\sum_{k=1}^{\infty}
  \bigl\{\rho(k+t)+\rho(k-t)-2\rho(k)\rho(t)\bigr\}^2
  \enum
}
for the variance of the normalized autocorrelation
function $\rho(t)=\Gamma(t)/\Gamma(0)$.

\subsection E.3 Practical procedure

Equations (E.6) and (E.10) can be evaluated by substituting
$\rhobar(t)=\Gbar(t)/\Gbar(0)$ for the normalized autocorrelation function
and truncating the sums at some sufficiently large time lag.
In the case of the variance of the autocorrelation function,
\equation{
  \left\langle\delta\rho(t)^2\right\rangle\simeq
  {1\over N}\sum_{k=1}^{t+\Lambda}
  \bigl\{\rhobar(k+t)+\rhobar(k-t)-2\rhobar(k)\rhobar(t)\bigr\}^2,
  \enum
}
the choice of the cutoff
$\Lambda$ is not critical and values of $\Lambda\geq100$ gave 
consistent results for all error bands plotted in fig.~6.

The integrated autocorrelation time may then be determined
through the sum
\equation{
  \tint={1\over2}+\sum_{t=1}^W\rhobar(t),
  \enum
}
in which the summation window 
$W$ is set to the first time lag $t$ where
\equation{
  \rhobar(t)\leq\left\langle\delta\rho(t)^2\right\rangle^{1/2}.
  \enum
}
Other criteria [\ref{UlliError},\ref{MadrasSokal}]
could be applied at this point and may be preferable
when the statistics allows
for an optimization of the summation window.
The Madras--Sokal formula
\equation{
  \left\langle\delta\tint^2\right\rangle\simeq
  {4W+2\over N}\tint^2
  \enum
}
may finally be used to estimate the statistical error of 
the autocorrelation time.

\beginbibliography


\bibitem{SesamTXLauto}
Th. Lippert et al. (SESAM and T$\chi$L collab.),
Nucl. Phys. B (Proc. Suppl.) 60A (1998) 311


\bibitem{SesamBig}
N. Eicker et al. (SESAM collab.),
Phys. Rev. D59 (1999) 014509


\bibitem{SesamTxLrzero}
G. S. Bali et al. (SESAM and T$\chi$L collab.),
Phys. Rev. D62 (2000) 054503


\bibitem{SesamTxLQuark}
N. Eicker, Th. Lippert, B. Orth, K. Schilling,
Nucl. Phys. B (Proc. Suppl.) 106 (2002) 209


\bibitem{GRALsize}
B. Orth, Th. Lippert, K. Schilling,
Nucl. Phys. B (Proc. Suppl.) 129 (2004) 173


\bibitem{Orth}
B. Orth, Finite-size effects in lattice QCD with dynamical Wilson
fermions, PhD thesis WUB-DIS 2004-03  (University of Wuppertal, Germany, 2004)


\bibitem{UKQCDbig}
C. R. Allton et al. (UKQCD collab.),
Phys. Rev. D60 (1999) 034507;
{\it ibid.} D65 (2002) 054502


\bibitem{UKQCDlight}
C. R. Allton et al. (UKQCD collab.),
Phys. Rev. D70 (2004) 014501


\bibitem{CPPACSbig}
A. Ali Khan et al. (CP-PACS collab.),
Phys. Rev. Lett. 85 (2000) 4674 [E: {\it ibid.} 90 (2003) 029902]; 
Phys. Rev. D65 (2002) 054505 [E: {\it ibid.} D67 (2003) 059901]


\bibitem{JLQCDlight}
S. Aoki et al. (JLQCD collab.)
Phys. Rev. D68 (2003) 054502


\bibitem{CPPACSsmall}
Y. Namekawa et al. (CP-PACS collab.),
Light hadron spectroscopy in two-flavor QCD with small sea quark masses,
arXiv: hep-lat/0404014


\bibitem{Saad}
Y. Saad, Iterative methods for sparse linear systems,
2nd ed. (SIAM, Philadelphia, 2003); see also
{\tt http://www-users.cs.umn.edu/\~{}saad/}


\bibitem{HMC}
S. Duane, A. D. Kennedy, B. J. Pendleton, D. Roweth,
Phys. Lett. B195 (1987) 216


\bibitem{ForcrandTakaishi}
Ph. de Forcrand, T. Takaishi,
Nucl. Phys. B (Proc. Suppl.) 53 (1997) 968

\bibitem{PeardonPreHMC}
M. J. Peardon. Accelerating the Hybrid Monte Carlo algorithm with ILU
preconditioning,
arXiv: hep-lat/0011080


\bibitem{Hasenbusch}
M. Hasenbusch,
Phys. Lett. B519 (2001) 177

\bibitem{HasenbuschJansen}
M. Hasenbusch, K. Jansen,
Nucl. Phys. B659 (2003) 299

\bibitem{DellaMorteEtAl}
M. Della Morte et al. (ALPHA collab.),
Comput. Phys. Commun. 156 (2003) 62


\bibitem{SchwarzAlgorithm}
M. L\"uscher,
J. High Energy Phys. 0305 (2003) 052

\bibitem{SchwarzSolver}
M. L\"uscher,
Comput. Phys. Commun. 156 (2004) 209


\bibitem{Wilson}
K. G. Wilson, Phys. Rev. D10 (1974) 2445

\bibitem{SW}
B. Sheikholeslami, R. Wohlert,
Nucl. Phys. B 259 (1985) 572

\bibitem{OaImp}
M. L\"uscher, S. Sint, R. Sommer, P. Weisz,
Nucl. Phys. B478 (1996) 365


\bibitem{Iwasaki}
Y. Iwasaki,
Renormalization group analysis of lattice theories and improved lattice action.
2. Four-dimensional non-abelian SU($N$) gauge model,
Tsukuba preprint UTHEP-118 (1983);
Nucl. Phys. B258 (1985) 141


\bibitem{LWaction}
M. L\"uscher, P. Weisz,
Commun. Math. Phys. 97 (1985) 59 [E: {\it ibid.} 98 (1985) 433];
Phys. Lett. B158 (1985) 250


\bibitem{SextonWeingarten}
J. C. Sexton, D. H. Weingarten,
Nucl. Phys. B380 (1992) 665

\bibitem{PeardonSexton}
M. J. Peardon, J. C. Sexton (TrinLat Collab.),
Nucl. Phys. B (Proc. Suppl.) 119 (2003) 985


\bibitem{SSOR}
S. Fischer et al.,
Comput. Phys. Commun. 98 (1996) 20;
Nucl. Phys. B (Proc. Suppl.) 53 (1997) 990


\bibitem{SommerScale}
R. Sommer,
Nucl. Phys. B411 (1994) 839


\bibitem{BiCGstabI}
H. A. van der Vorst, 
SIAM J. Sci. Stat. Comput. 13 (1992) 631

\bibitem{BiCGstabII}
A. Frommer, V. Hannemann, B. N\"ockel, T. Lippert, K. Schilling,
Int. J. Mod. Phys. C5 (1994) 1073


\bibitem{ReplayTrick}
M. L\"uscher, R. Sommer,
in preparation


\bibitem{UlliError}
U. Wolff,
Comput. Phys. Commun. 156 (2004) 143
[E: hep-lat/0306017 v3]

\bibitem{MadrasSokal}
N. Madras, A. D. Sokal,
J. Stat. Phys. 50 (1988) 109




\bibitem{NumMethods}
L. Giusti, C. Hoelbling, M. L\"uscher, H. Wittig,
Comput. Phys. Commun. 153 (2003) 31

\endbibliography

\bye